\documentclass{amsart}

\usepackage{amsfonts,amsmath,amsthm,mathrsfs,dsfont,mathtools}
\usepackage{verbatim,graphicx,psfrag}
\usepackage{euscript}
\usepackage{upgreek}
\usepackage{wrapfig}
\usepackage{tikz-cd,braids}
\usepackage{diagbox}
\usepackage{hyperref}
\newcommand{\0}{\mathbf{0}}

\newcommand{\ba}{\mathbf{a}}
\newcommand{\bb}{\mathbf{b}}
\newcommand{\bc}{\mathbf{c}}
\newcommand{\bd}{\mathbf{d}}

\newcommand{\bm}{\mathbf{m}}

\newcommand{\bp}{\mathbf{p}}

\newcommand{\bx}{\mathbf{x}}
\newcommand{\by}{\mathbf{y}}
\newcommand{\bz}{\mathbf{z}}
\newcommand{\bomega}{\boldsymbol{\omega}}

\newcommand{\cH}{\mathcal{H}}
\newcommand{\cM}{\mathcal{M}}

\newcommand{\gh}{\mathfrak{h}}

\newcommand{\ha}{\hat{a}}
\newcommand{\hb}{\hat{b}}
\newcommand{\hc}{\hat{c}}

\newcommand{\hp}{\hat{p}}
\newcommand{\hx}{\hat{x}}

\newcommand{\sU}{\mathrm{U}}

\newcommand{\C}{\mathbb{C}}

\newcommand{\N}{\mathbb{N}}
\newcommand{\bbP}{\mathbb{P}}

\renewcommand{\S}{\mathbb{S}}
\newcommand{\R}{\mathbb{R}}
\newcommand{\Z}{\mathbb{Z}}

\newcommand\1{{\ensuremath {\mathds 1} }}
\newcommand{\norm}[1]{\left\lVert #1 \right\rVert}

\DeclareMathOperator{\proj}{\mathrm{proj}}
\DeclareMathOperator{\spec}{\mathrm{spec}}

\definecolor{darkred}{cmyk}{0,1,1,0.4}
\definecolor{darkblue}{rgb}{0,0,60}
\newcommand{\dred}{\textcolor{darkred}}
\newcommand{\dblue}{\textcolor{darkblue}}

\DeclarePairedDelimiterX\braket[2]{\langle}{\rangle}{#1 \delimsize\vert #2}

\def\Z{\mathbb{Z}}

\def\R{\mathbb{R}}

\newcounter{thm}
\newcounter{ex}
\newcounter{re}
\newtheorem{Theorem}[thm]{Theorem}

\newtheorem{Proposition}[thm]{Proposition}
\newtheorem{example}[thm]{Example}
\newtheorem{remark}[thm]{Remark}
\newtheorem{Definition}[thm]{Definition}

\title[Six blinds, measurements and the elephant]{The six blinds and the elephant or an interdisciplinary selection of measurement features}

\author{Ask Ellingsen}
\author{Douglas Lundholm}
\address{A.E. \& D.L.: Department of Mathematics, Uppsala University, Box 480, SE-751 06, Uppsala, Sweden}
\email{ask.ellingsen@math.uu.se}
\email{douglas.lundholm@math.uu.se}

\author{Jean-Pierre Magnot}
\address{J.-P.M.: Univ. Angers, CNRS, LAREMA, SFR MATHSTIC, F-49000 Angers, France
	\\ and \\  Lyc\'ee Jeanne d'Arc \\ Avenue de Grande Bretagne, \\ 63000 Clermont-Ferrand, France}
\email{magnot@math.cnrs.fr}
	
\begin{document}
	
	\begin{abstract}
	We propose here selected actual features of measurement problems based on our concerns in our respective fields of research. Their technical similarity in apparently disconnected fields motivate this common communication. Problems of coherence and consistency, correlation, randomness and uncertainty are exposed in various fields including physics, decision theory and game theory, while the underlying mathematical structures are very similar.    	  
	\end{abstract}
	
	\maketitle
	
	\textit{Keywords:} measurement, gauge theory, holonomy, quantum statistics, anyons, pairwise comparison, impossible figures, contextuality, telepathy, free will, computation
	
	\smallskip
	
	\smallskip
	
 


	\section*{Introduction}
	\begin{wrapfigure}{r}{0.5\textwidth}
		\centering
		\includegraphics[scale=0.26]{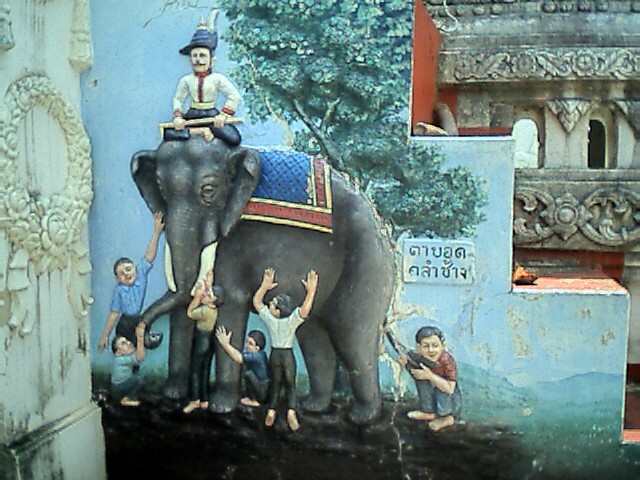}
		\caption{Wall relief in North-East Thailand}
	\end{wrapfigure}
The six blinds and the elephant is a famous tale 
present in Jainism, Hinduism, Buddhism and Sufism, that accounts how six blind men 
fall into confusion 
when sharing their own description of an elephant to each other, after touching the elephant one from one side, one from  another side. 

One of the lessons of this parable is the difficulty of measuring the whole reality from partial measurements, and intends to recall that men need to remain humble whatever they can know or discover, the full realm being beautifully complex and unreachable. 
Here we intend to describe a number of similar situations,
from physics to economics to cartography, 
passing by quantum information and the art of 
perspective drawing,
all approached using mathematical tools of the same kind.
Just like the blinds, 
in trying to integrate partial information
we stumble upon a number of logical concerns,
but arising from such
different approaches that sometimes even the vocabularies used to describe
the problems are different. 
A deep, unified, and coherent picture seems hidden to our blind eyes. Contemplating these problems of hidden complexity of the existing ``real'' world, we reach the limits of Laplace's 
deterministic worldview, 
and also the limits of classical logic 
as a common language for the description of mathematical objects, 
as suggested in e.g. \cite{Bu2023,Wa2023}.
There are natural epistemological questions pending to this 
philosophical problem,
see e.g. \cite{Mu2023},
and observations of this type seem
to be ever rising in modern physics, as 
illustrated by the following notations quoted in \cite{Desmet-22} and \cite{St2019}:

\vskip 6pt
\noindent
Albert Einstein: ``You can never solve a problem on the level on which it was created.''
\vskip 6pt
\noindent
Eugene Paul Wigner: ``Mathematical concepts turn up to entirely unexpected connections. Moreover, they often permit an unexpectedly close and accurate description of the phenomena in these connections. Secondly, just because of this circumstance, and because we do not understand the reasons of their usefulness, we cannot know whether a theory formulated in terms of mathematical concepts is uniquely appropriate.''
\vskip 6pt  
\noindent
Ren\'e Thom: 
``That portion of reality, which can be well described by laws which permit calculations, is extremely limited. [...] 
In my view, any groundbreaking theoretical progress is a consequence of the ability of the researchers to `get into the skin of things', to be able to empathically identify with any entity of the exterior reality. And this kind of identification transforms an objective phenomenon into a sort of concrete mental experience.''
\smallskip

However, our communication does not intend to be epistemological. It would be more accurate to say that we humbly present an overview of the (yet, epistemological) problems that we have perceived in our technical investigations on mathematical tools in various areas of applied mathematics and physics than to pretend to give a deep insight in the problems that we raise, first because the very wide scope of knowledge under consideration.    
By invoking mathematical tools, we have here to be precise. What is a mathematical tool? A number, a probability or an (often, self-adjoint) operator? These are often the superficial aspects that a common observer can see, but we include also models, algebra, analysis and geometric structures, and even more fundamentally, logic. All these so-called tools are mathematics, the science developed by human beings in order to build models and understand, choose or make predictions. This is what we understand here as measurements beyond the ``classical'' measurements, and that is the core of our paper.

Indeed, our work considers measurements and comparisons, from the viewpoint of information captured by an observer in a localized place and with respect to a particular context. Therefore, we describe our own understanding of the notions of correlation, locality, coherence and inconsistency, as well as randomness in measurements. This exposition is based on the mathematical concepts of non-triviality of fiber bundles and 
connections, as well as random variables, with applications in anyon gas physics, loop quantum gravity, pairwise comparisons in decision theory, and illustrated using ideas of game theory, free will, and quantum pseudo-telepathy.

As in our starting parable, we propose this exposition in one plus six sections, and we highlight a few open questions in a final outlook section. 
 
	\section{A selected overview on measurements and comparisons}
	\subsection{Physics as information and correlations}
 
Physics intends to describe the ``real'' world by means of repeatable experiments and measurements. 
To achieve this, it typically makes a logical division of the world into {\bf systems/subsystems and observers}.
The first ones are measured while the second ones receive the measurements on the first ones.
The measurements and the reception of the measurements by the observer involve {\bf information}. Therefore, the couple (system, observer) is itself a system in which information is an internal relationship. The information on a system accessible by an observer is in general called \textbf{observable}. 

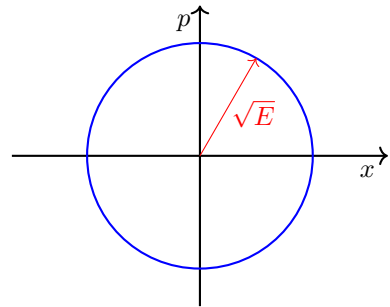
\begin{wrapfigure}{r}{0.5\textwidth} 
    \centering
		\begin{tikzpicture}
			\draw [arrows=->,thick] (-2.5,0) -- (2.5,0);
			\draw [arrows=->,thick] (0,-2) -- (0,2);
			\draw [blue,thick] (0,0) ellipse (1.5 and 1.5);
			\draw [arrows=->,red] (0,0) -- (0.75,1.299);
			\node [below right] at (2,0) {$x$};
			\node [below left] at (0,2) {$p$};
			\node [below right,red] at (0.3,0.866) {$\sqrt{E}$};
		\end{tikzpicture}
    \caption{Correlations between observables $p$, $x$ and $E$ in the description of the harmonic oscillator.}
    \label{fig:correlations}
\end{wrapfigure}

In the classical view of physics (up until the early 1900's),
observables are considered the
aspects of a system that can be measured and have well-defined values as {\bf properties of reality} (sometimes called states of the reality, but this vocabulary may be misleading), for example
$\mathbf{x} \in \R^3$ is the position of a particle with respect to a laboratory reference frame, 
$\mathbf{p} \in \R^3$ is its momentum, 
$E \in \R$ is its energy, 
and $t \in \R$ 
is time measured by a clock.
Physics tells us that these quantities may be linked together by 
\emph{relations} which govern their common dependence. These are {\bf correlations} between observables. For example, for a particle described by classical mechanics in one dimension with spatial coordinate $x$ and momentum $p$, the energy could be given by the simple relationship
$E = p^2 + x^2$, describing (in dimensionless units) the periodic motion in a harmonic oscillator potential centered at $x=0$.
Therefore, at constant energy, the observables $x$ and $p$ are correlated as the points of a circle of radius $\sqrt{E}$ in the $x$-$p$-plane (cf.\ Fig.~\ref{fig:correlations}). 
We reach here the (basic) set-theoretical notion of relation as a subset of a cartesian product.
Thus, from this classical perspective our view of physics is mainly about the study of such relationships between various observables.

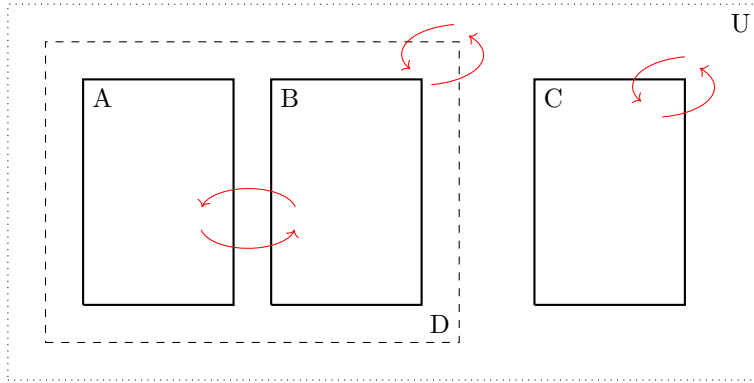
\begin{figure}
	\begin{tikzpicture}
		\draw [dotted] (-5,-1) -- (5,-1) -- (5,4) -- (-5,4) -- (-5,-1);
		\draw [dashed] (-4.5,-0.5) -- (1,-0.5) -- (1,3.5) -- (-4.5,3.5) -- (-4.5,-0.5);
		\draw [thick] (-4,0) -- (-2,0) -- (-2,3) -- (-4,3) -- (-4,0);
		\draw [thick] (-1.5,0) -- (0.5,0) -- (0.5,3) -- (-1.5,3) -- (-1.5,0);
		\draw [thick] (2,0) -- (4,0) -- (4,3) -- (2,3) -- (2,0);
		\node [below right] at (-4,3) {A};
		\node [below right] at (-1.5,3) {B};
		\node [below right] at (2,3) {C};
		\node [above left] at (1,-0.5) {D};
		\node [below left] at (5,4) {U};
		\draw [red] (-2.43,1.0) [xscale=2.1,->] arc(-170:-10:.30);
		\draw [red] (-1.18,1.3) [xscale=2.1,->] arc(10:170:.30);
		\draw [red] (0.93,3.73) [xscale=2.1,->] arc(100:210:.40);
		\draw [red] (0.63,2.93) [xscale=2.1,->] arc(-80:40:.40);
		\draw [red] (4.0,3.3) [xscale=2.1,->] arc(100:210:.40);
		\draw [red] (3.7,2.5) [xscale=2.1,->] arc(-80:40:.40);
	\end{tikzpicture}
    \caption{Correlations between various subsystems of U.}
    \label{fig:subsystems}
\end{figure}

The limit of this picture is reached when the conditions of measurement change the observed quantities (for example, one cannot meet wild bisons in Bialowieza every day). 
Indeed, in the course of the previous century with the advent of quantum mechanics, the distinction between the observed subsystem and the observer subsystem has become less sharp. One may argue that all subsystems of the universe are quantum and should be treated on an equal footing \cite{Rovelli-96},
and instead of the correlations of one subsystem's observables with respect to an {\bf objective observer}, the role of correlations in information (or knowledge) between subsystems becomes the focus of physics --- a move towards {\bf subjectivity}.
An observable could thus typically concern the information that subsystem A has on subsystem B, or that B has on A, as viewed from a supersystem D, or the information that D has on A or on another supersystem U, or that U has on yet another subsystem C, etc. (cf. Fig.~\ref{fig:subsystems}). 
It is rather the intricate network of subsystems and the various relations and dependencies which constrain or facilitate correlation in information,
for example the geometry and topology of ensembles of subsystems, that define (model) physical reality and its properties.

\subsection{Locality and the gauge problem}
	The gauge problem starts with locality.
	
	\begin{wrapfigure}{r}{0.5\textwidth}
		\centering
		\includegraphics[scale=0.5]{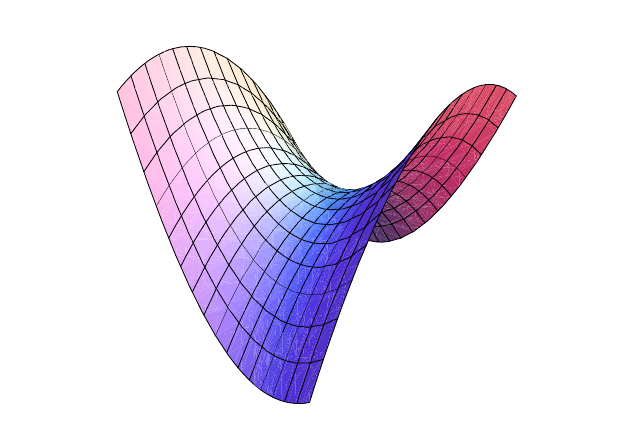}
		\caption{\newline Curved saddle surface}
        		\label{fig:saddle}
	\end{wrapfigure}
	
 \noindent
	 Many measurements are local. For example, in a saddle surface (cf. Fig.~\ref{fig:saddle}), the area, the orientation of the normal vector are highly dependent of the chosen point of the surface.
	
	\begin{wrapfigure}{r}{0.5\textwidth}
		\centering
		\includegraphics[scale=0.15]{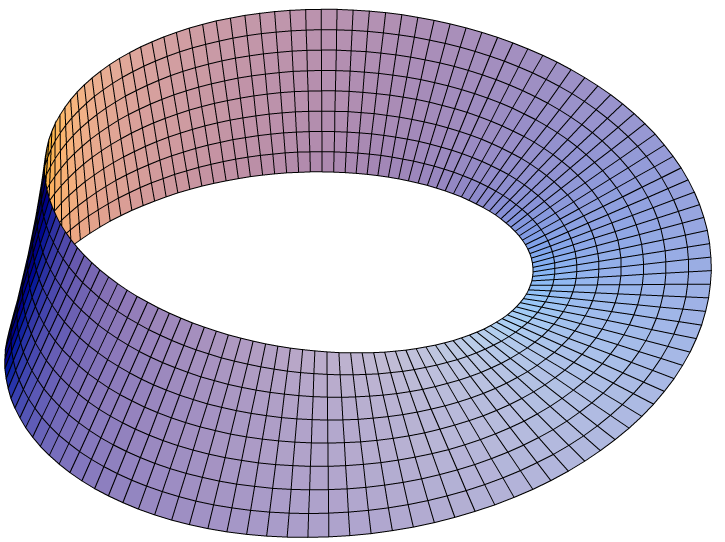}
		\caption{\newline M\"obius strip}
        		\label{fig:mobius}
	\end{wrapfigure}
 
	Even more, considering the M\"obius strip (cf. Fig~\ref{fig:mobius}), a loop along the strip finds reverse orientation. Therefore, a measurement or an evaluation depends on the place of the evaluation. 
	
	If the evaluations can be mapped smoothly by \emph{local coordinates} identified as an open subset of a topological vector space, we work on a \emph{manifold} $M$ and the local evaluations take values in a \emph{fiber bundle} $E$ over $M$. The fiber bundle is equipped with a projection $\pi : E \rightarrow M$ for which each subset $\pi^{-1}(x),$ for $x \in M,$ is isomorphic to a \emph{typical fiber} $F,$ and with a system of local trivializations over $M,$ and there exists a (non-linear) frame bundle $\mathrm{Fr}(E)$ made of fiberwise isomorphisms between $F$ and a fiber $\pi^{-1}(x).$ $\mathrm{Fr}(E)$ is a principal bundle with structure group $\mathrm{Diff}(F)$ (diffeomorphisms), and if $F$ possesses additional structures, the structure group can be reduced. For example, when $E$ is a \emph{vector} bundle, that is, when the fibers are equipped with a (finite dimensional) vector space structure (intuitively speaking), then so is $F$ and we can replace $\mathrm{Diff}(F)$ by the general linear group $\mathrm{GL}(F),$ which describes the (classical) bundle of frames of a manifold when $E = TM.$ We refer to \cite{Hus66,KN63-69,KM,KMS} as classical references on this subject, both for finite-dimensional and infinite-dimensional cases.  
	\begin{remark}
		There also exist a more generalized notion of fiber pseudo-bundles, where there may exist no local trivialization, and where the fiber may change. This happens for example in generalized definitions of tangent spaces for algebraic varieties, but we omit these problems in our overview.
		\end{remark}
Therefore, even in simple examples, it is difficult to choose a frame or another for measurements. This is the gauge problem, where seemingly no frame is preferable to another, except for physical motivations. 

For what follows, we need more advanced notions of differential geometry such as differential forms, distributions among others, that we cannot sketch here for length considerations. We refer to the already cited textbooks, as well, as to the very classical and quite complete \cite{KN63-69}.  In order to define a local or a global gauge, a common tool consists of a \emph{connection}, which can be identified as a differential form $\theta \in \Omega^1(\mathrm{Fr}(E),\mathfrak{gl}(F))$ in the case of a vector bundle $E$, where $\mathfrak{gl}(F)$ is the Lie algebra of the linear group $\mathrm{GL}(F),$ which is equivariant with respect to the right action of $\mathrm{GL}(F)$ on $\mathrm{Fr}(E)$ (the action is by composition on the right), and which defines an isomorphism between $V_p \mathrm{Fr}(E) = T_pE \cap D_p\pi^{-1}(0)$ and $F,$ and a horizontal distribution $HE$ which splits the tangent space $TE$ into $TE = HE \oplus VE$, where $VE = \mathrm{Ker}\, \pi.$

One can check that these two definitions are equivalent for finite dimensional vector bundles (see e.g. \cite{KN63-69}) and a connection always defines a local slice in $E$ over any star-shaped domain $D \subset M$ by integration of the horizontal distribution along the radii of the domain $D$; see e.g.\ \cite{Lich} for a comprehensive exposition. 
\begin{wrapfigure}{r}{0.5\textwidth}
	\centering
	\includegraphics[scale=0.2]{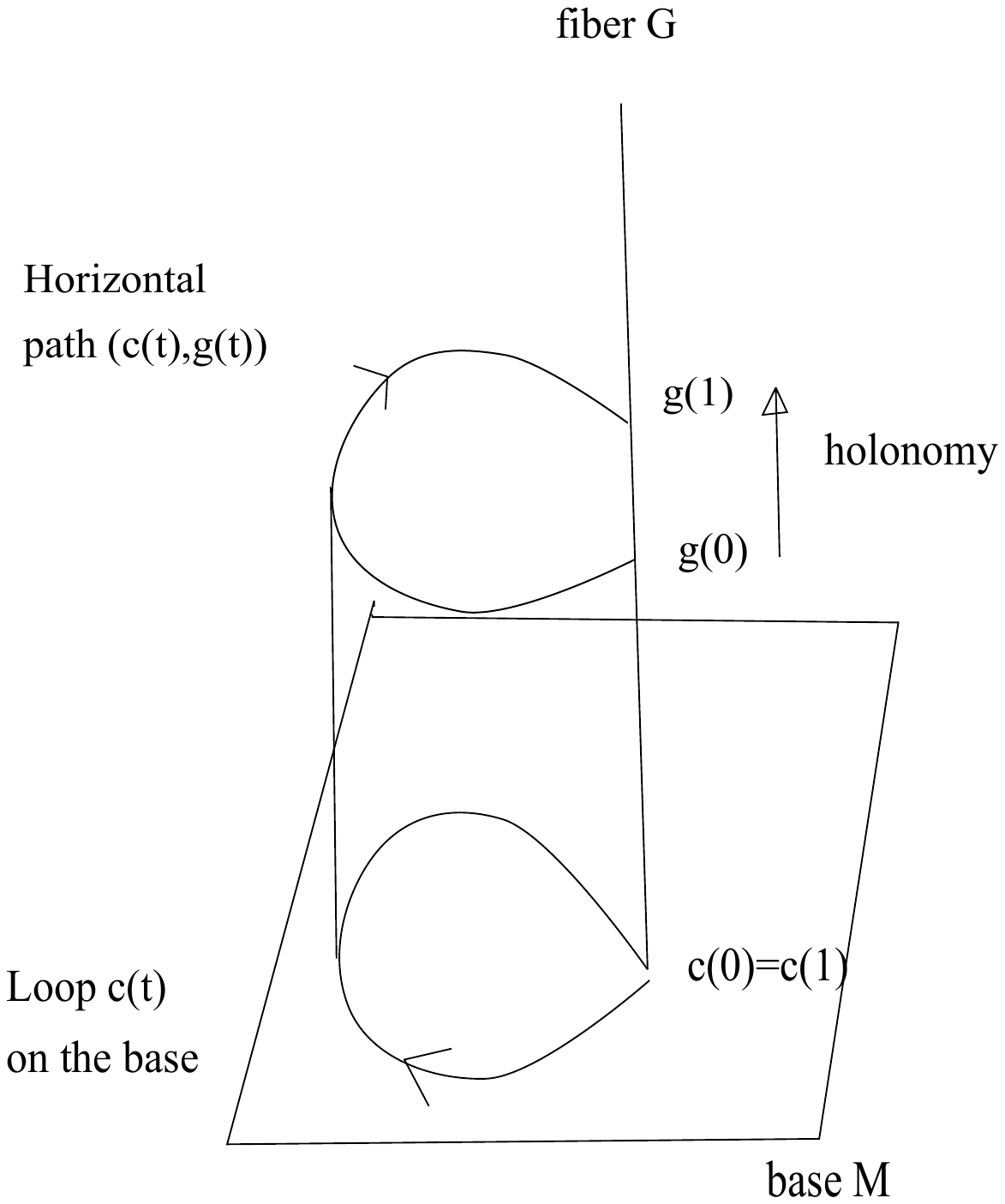}
	\caption{Holonomy of a non-flat connection.}
        \label{fig:holo}
\end{wrapfigure} 
When the connection defines a global section, the vector bundle $E$ is then identified with $M \times F$ via the choice of the global gauge.

\begin{example}
	When measuring altitude on earth, the gauge is the altitude of the surface of the oceans. This measure is performed with respect to radial axes passing at the center of the earth, approximized as a sphere. This means that one is able to define "levels of altitude" all around the earth, even when the base point of the radial axis is not in an ocean. This then defines a global gauge, motivated by physical considerations.
\end{example}

When the horizontal distribution $HE$ does not integrate, at least locally, to a slice from a star-shaped domain $D$ to $\pi^{-1}(D),$ it implies that there exists a loop on $D$ whose horizontal lift is not a loop but a path. This horizontal path $Hc$ has a non-trivial holonomy $h$ if $Hc(1)=Hc(0).h$ 
(and $h \neq \1$ depends on the base point $Hc(0)$ if $G$ is non-abelian).

This is characterized by the following property:

\begin{Proposition}
	Let $D$ be a star-shaped domain in $M$ and let $D \times G$ be a local trivialization of $E$ over $D.$ The horizontal lift $(c(t),g(t))$ of a loop $c$ on $D$ satisfies the logarithmic equation $\partial_t g(t). g^{-1}(t) = \theta(\partial_t c(t)).$ Then 
    $d\theta + [\theta,\theta] = 0$ 
    (the curvature is zero)
    if and only if the horizontal lift of any loop is a loop. 
\end{Proposition}

\subsection{On random measurements}
Even from a naive viewpoint, it is clear that a measurement can carry limited precision or can be endowed with errors. Therefore, what is called deterministic measurement can only be an ideal viewpoint, or a simplified version of a more complex realm. For this reason, a measurement can be modeled as a random variable which intends to model the unknown parameters of the experience and the dependence of the results on them.

Let us consider two concrete examples:
\begin{itemize}
	\item A measurement of length is done by a school pupil who performs his work with care. He is measuring 42 millimeters and he cannot make better measure in terms of precision. This lack of precision suggests that the measure is random and its probability law is uniform, with support in an interval with one millimeter length. 
	\item The measurement of the same length, depending on the person who performs the measure, can slightly differ. In that case, it appears that the expectation value of the random variable remains 42 millimeters, but it appears that the probability law can be approximated by a Gaussian law.    
\end{itemize}

These are basic motivations for statistical physics. However, a much deeper reason comes from Heisenberg's uncertainty principle which (roughly speaking) asserts that we cannot measure positions at a better precision than Planck's constant. But the Heisenberg's principle cannot be reduced to this statement which has its corresponding property in the theory of wavelets \cite{Mey92}, and will be precised a little later.  

\subsection{Physics as simultaneous/coherent information}

    While classical mechanics deals with both kinematical and dynamical aspects of correlations between observables, 
    quantum mechanics deals with the logical compatibility of observables and of the information we may have about them.
    Typically, and in brief (see e.g. \cite{Lundholm-19}), 
    one may view {\bf quantization} as a choice of representation of a {\bf Lie algebra} of classical observables into the algebra of linear operators on some Hilbert space.
    That is, we have made a selection of a set of \emph{relevant} real-valued observables, which we denote $a,b,c,\ldots$, and which are modeled as self-adjoint linear operators
    $\ha,\hb,\hc,\ldots$ acting jointly\footnote{Here we will ignore any issues of unbounded operators and even consider finite dimensions.} on a Hilbert space 
    $(\cH, \langle \cdot,\cdot \rangle)$.
    Linear combinations of observables $\alpha a + \beta b$ are mapped to
    linear combinations of operators $\alpha \ha + \beta \hb$,
    and, importantly, we also have a Lie bracket
    \begin{equation}\label{eq:commutator}
        [\ha,\hb] = \ha\hb - \hb\ha = i\hc,
    \end{equation}
    which typically comes induced from a classical Poisson bracket $\{a,b\} = c$.
    These mathematical structures allow then to make the following physical interpretations:
    \begin{itemize}
    \item The {\bf spectrum} $\spec\ha = \{\lambda\}$ of the self-adjoint operator $\ha$ corresponding to an observable $a$ are the possible values that $a$ can take upon measurement.
    
    \item The {\bf spectral resolution} of $\ha$ in terms of projections onto its eigenspaces,
    $$\ha = \sum_{\lambda \in \spec\ha} \lambda \proj_{\cH_{a = \lambda}},$$
    expresses the information that can be known about $a$,
    i.e. the totality of information about $a$ that is possible to extract from the system (including correlations with other observables).

    \item A {\bf state} in $\cH$ is a nonzero vector $\Psi \in \cH$
    and represents the actual information known about the system,
    encoded in the way it projects into the spectral subspaces,
    $$\cH = \bigoplus_{\lambda \in \spec\ha} \cH_{a = \lambda},$$
    i.e. it represents our current, subjective `reality' about $a$.

    \item The {\bf expectation}
    $$
        \frac{\langle \Psi,\ha\Psi \rangle}{\norm{\Psi}^2} 
        = \sum_{\lambda \in \spec\ha} \lambda \frac{\norm{\proj_{\cH_{a=\lambda}}\Psi}^2}{\norm{\Psi}^2}
    $$
    represents the probability distribution for the values of $a$,
    given that we know the state $\Psi$,
    where the relative norm 
    $\norm{\proj_{\cH_{a=\lambda}}\Psi}^2/\norm{\Psi}^2$
    of the component of $\Psi$ in the spectral subspace $\cH_{a=\lambda}$
    gives the probability that $a = \lambda$.
    The components $\proj_{\cH_{a=\lambda}}\Psi$ are known as {\bf probability amplitudes}.

    \item The {\bf commutator} \eqref{eq:commutator}
    represents the obstacle (the obstruction) to simultaneous information on $\ha$ and $\hb$. 
    We call two observables $a,b$ {\bf incommensurate} if their commutator is nonzero.
    
    \end{itemize}

\subsection{Heisenberg's uncertainty principle}
 
    The {\bf uncertainty principle} is fundamental to quantum mechanics and illustrates this inability to obtain simultaneous information on incommensurate observables. 
    For example, let
    $$
	A = \left( \begin{array}{cc} 1&0\\0&-1 \end{array}\right) \ \hbox{and} \ 
	B = \left( \begin{array}{cc} 0&1\\1&0 \end{array}\right).
    $$
    be self-adjoint matrices acting on the Hilbert space $\cH = \C^2$,
    and may thus be thought of as quantum observables 
    (they are actually Pauli matrices or spin operators acting on a quantum bit).
    They satisfy $A^2 = B^2 = \1$ and have the same spectrum $\{-1,+1\}$.
    However, we have that 
    $AB = -BA$, 
    which implies that $[A,B] \neq 0$ but even more, in this precise (extremal) case, this 
    means that obtaining knowledge of one destroys knowledge of the other.
    This is encoded in their spectral resolutions, namely their respective eigenspace decompositions
    $$
        \cH = \C^2 = \C(1,0) \oplus \C(0,1) = \C(1,1) \oplus \C(1,-1)
    $$
    are incompatible.
    For example, if we know with certainty that $B$ has the value $+1$, then $\Psi \propto (1,1)$, and thus we know only that $A$ has the value $+1$ or $-1$ with $50\%:50\%$ probability.
	
    Heisenberg considered a continuous version of the above, namely the position and momentum observables $x \in \R$ and $p \in \R$,
    which are {\bf canonically conjugate}, meaning that their commutator is the identity:
    $$
        \hx\hp - \hp\hx = i\1.
    $$
    Quantization in this case boils down to the 
    {\bf Schr\"odinger representation}\footnote{It 
    is necessary to represent this algebra using unbounded operators in infinite dimensions, so we now allow for this generality, with e.g.\ the formal eigenprojections $\proj_{\cH_{x=x_0}} = \delta_{x_0} \langle\delta_{x_0}, \cdot \rangle$.} 
    $$
	\hx\Psi(x) := x\Psi(x), \qquad
        \hp\Psi(x) := -i\Psi'(x),
    $$
    $$
	x(-i\Psi'(x)) - (-id/dx)(x\Psi(x)) = i\Psi(x),
    $$
    on the Lebesgue space or fiber integral
    $$
        \cH = L^2(\R;\gh) = \int^\oplus_{\R} \gh, 
    $$
    i.e.\ the square-integrable measurable functions
    $\Psi\colon \R \to \gh$,
    where the fiber Hilbert space $\gh$ may admit representations of any additional observables.
    The expectation of position is
    $$
	\frac{\langle \Psi,\hx\Psi \rangle}{\norm{\Psi}^2} 
		= \int_{-\infty}^\infty x \frac{|\Psi(x)|_\gh^2}{\norm{\Psi}^2} \,dx,
    $$
    while that of the momentum is obtained using the Fourier transform.
    Due to their incommensurability, we cannot simultaneously localize knowledge
    of $\hx$ and $\hp$ arbitrarily sharply to any combination $(x,p) \in \R^2$, however we can have states that distribute over a range of joint values, such as the
    eigenstates of the energy observable $\hat{E} := \hp^2 + \hx^2$ of the quantum harmonic oscillator.

\section{Twisting gives us quantum statistics, anyons, and stability}
	
    In this section, we consider the interplay of uncertainty between different observables and the role of twisting to produce another very tangible effect of quantum mechanics: 
    the {\bf Pauli exclusion principle} that is responsible for the periodic table of the elements, the extensivity of matter, and the stability of very large systems such as planets and stars against electromagnetic or gravitational collapse.

    \subsection{Twisting on a circle}

    On the circle $\S^1$ we can use {\it locally} the same (flat) quantization as on the real line, i.e.\ we have canonically conjugate position and momentum operators
	$$
		\hat{\varphi}\Psi(\varphi) = \varphi\Psi(\varphi) \ \text{for} \ \varphi \in (0,2\pi), \qquad
		\hp_\varphi = -i \frac{d}{d\varphi},
	$$
    acting on the fibered Hilbert space $\cH = L^2([0,2\pi];\gh)$ over the angle $\varphi$.
    The {\bf identification} of positions $\varphi=0$ and $\varphi=2\pi$ 
    requires also an identification
    in the fiber $\gh$, i.e.\ a {\it global} (topological) boundary condition (b.c.), such as:
	$$
		\Psi(2\pi) = T\Psi(0), \quad T \in \sU(\gh).
	$$
    If the fiber is $\gh = \C$, the unitary operator $T$ is simply a {\bf twist} by an angle $\theta \in [0,2\pi)$:
	$$
		\Psi(2\pi) = e^{i\theta}\Psi(0),
	$$
    which allows to decompose our space
    $\cH = L^2(\S^1) = \bigoplus_{n \in \Z} \gh_n$
    in terms of the twisted Fourier series $f_n(\varphi) = e^{i(n+\theta/(2\pi))\varphi}$,
    with $\gh_n = \C f_n$ eigenspaces of $\hp_\varphi$ subject to these b.c.
    The result will be to shift the spectrum of the energy operator by an amount quadratic in the amount of twist (note the periodicity):	
	$$
		\hat{E} 
		:= \hp_\varphi^2
		= \bigoplus_{n \in \Z} (n+\theta/(2\pi))^2 \1_{\gh_n}
		\ge \min_{n \in \Z} |n+\theta/(2\pi)|^2.
	$$
    An interpretation of $T$ or $e^{i\theta}$ is the holonomy in the fiber $\gh$ associated with a simple, positively oriented, loop around the configuration space $\S^1$.
    For 
    $T \neq \1$
    the space $\cH$ consistent with the twist will have to be interpreted as sections of a fiber bundle over $\S^1$ or suitably equivariant functions on its covering space $\R \ni \varphi$.
    The locally flat such Hilbert vector bundles are in one-to-one correspondence with the global holonomy around the loop,
    parametrized (modulo unitary conjugation) by $T$.

    \subsection{Twisting in the plane}

    Consider now at least {\bf two} commensurate observables $(x_1,x_2) \in \R^2$
    with their conjugates $(p_1,p_2)$ and a correlating energy observable, ex.
	$$
		\hat{E} = \hp_1^2 + \hp_2^2 = -\frac{\partial^2}{\partial x_1^2} -\frac{\partial^2}{\partial x_2^2},
	$$
    the Laplacian.
    In polar coordinates $(r,\varphi) \in \R_+ \times [0,2\pi)$:
	$$
		\hat{E} = -\frac{\partial^2}{\partial r^2} - \frac{1}{r}\frac{\partial}{\partial r} - \frac{1}{r^2} \frac{\partial^2}{\partial \varphi^2}.
	$$
    Thus, in this description we make a fibration of the plane by circles, 
    $$\cH = L^2(\R^2) = L^2(\R_+;\gh) = L^2(\R_+) \otimes \gh, \qquad \gh = L^2(\S^1)$$
    Again, we may choose to represent
    the observable $\hp_\varphi = -i \partial/\partial\varphi$ on 
    the angular fiber
    $\gh = \bigoplus_{n \in \Z} \gh_n$
    with a {\bf twist} (uniform\footnote{Twists that depend on $r$ can also be prescribed, however, we avoid it in this example for reasons of scale covariance.} in $r$): 
    $\Psi(r,2\pi) = e^{i\theta}\Psi(r,0)$,
    so that
	$$
		\hat{E} = \bigoplus_{n \in \Z} \left( -\frac{\partial^2}{\partial r^2} 
		- \frac{1}{r}\frac{\partial}{\partial r}
		+ \frac{1}{r^2} (n+\theta/(2\pi))^2 \right) \otimes \1_{\gh_n}.
	$$
    Finally, due to the singularity at $r=0$, self-adjointness of the observable $\hat{E}$ requires also a choice of boundary condition (or domain) for any allowed states $\Psi \in \cH$,
    such as\footnote{This is the most natural one, the Friedrichs extension,
    corresponding to free particles, however other extensions may be chosen,
    describing point-interacting particles. See \cite{Lundholm-23} for an overview.} 
    $$
        \Psi \sim J_\alpha(r) e^{i\alpha\varphi} \sim r^\alpha e^{i\alpha\varphi} \ \text{as} \ r \to 0, 
        \qquad \text{where} \qquad
        \alpha = \min_{n \in \Z} |n+\theta/(2\pi)|,
    $$
    and $J_\alpha$ denotes the Bessel function of order $\alpha$.
    Thus, we find  
    as a consequence of the twist $\theta$ in the angular observable $\varphi$,
    that there is in $\Psi$ a damping dependence on the radial observable $r$, 
    and in fact a probability vortex around the point $r=0$, 
    $|\Psi(r,\varphi)|^2 \sim r^{2\alpha}$,
    arising due to their correlation via the domain of the energy observable $E = p_1^2 + p_2^2$ that also incorporates the uncertainty principle by the incommensurability of $x_j$ and $p_j$.
    One may refer to this vortex of reduced probability as an {\bf angular-momentum barrier}, and the example describes \emph{fractional} angular momentum $\alpha$.

\begin{figure}
	\begin{tikzpicture} 
		\node at (-5.7,0) {\scalebox{0.4}{\includegraphics{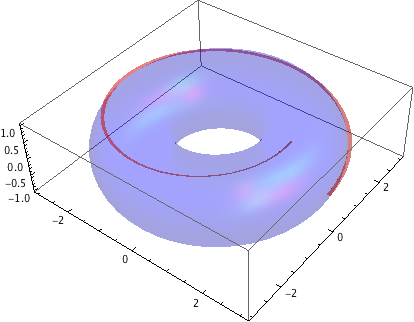}}};
		\node at (0,0){\scalebox{0.4}{\includegraphics{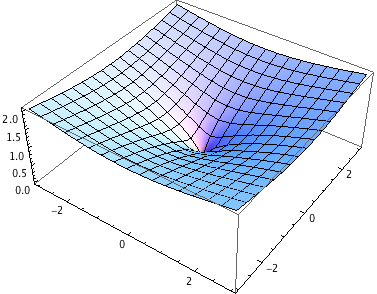}}};
		\node [above right] at (-4.3,1.5) {\small $\arg \Psi$};
		\node [above right] at (1.4,1.5) {\small $|\Psi|^2$};
	\end{tikzpicture}
    \caption{Twisting in the phase of a wave function may imply vorticity in probability, hence effective repulsion, hence exclusion.
    Here $\Psi(r,\varphi) \sim r^{\alpha} e^{i\alpha\varphi}$,
    where $\alpha = \theta/(2\pi) = 1/4$.}
    \label{fig:twist-exclusion}
\end{figure}

    \subsection{Twisting of identical particles}

    The quantum particles found in nature, such as electrons and photons, are indistinguishable to any measurement and in fact arrange themselves in accordance to being logically identical, by the laws of {\bf quantum statistics}.
    Since the 1920's we divide such identical particles into two families: {\bf bosons} and {\bf fermions}, obeying Bose-Einstein statistics resp.\ Fermi-Dirac statistics. The latter incorporates Pauli's exclusion principle: that fermions must occupy different one-body states, and implies that they prefer to avoid each other in space.
    More recently, intermediate quantum statistics in the plane and {\bf ``anyons''} were discovered, independently and from three different perspectives,
    by Leinaas \& Myrheim in 1977 in a geometric perspective using Schr\"odinger quantization on fiber bundles \cite{LeiMyr-77},
    by Goldin, Menikoff \& Sharp in 1980-'81 in an algebraic perspective using Heisenberg quantization of current algebras and diffeomorphism groups \cite{GolMenSha-80,GolMenSha-81},
    and by Wilczek in 1982 using a magnetic perspective \cite{Wilczek-82a,Wilczek-82b}.
    In fact, the example we considered above in the plane is directly applicable to these general quantum statistics in order to deduce their exclusion, by considering a pair of particles in their relative position and momentum.
    In the plane $\R^2$ the relative angle of two identical particles lives on the half-circle (or the circle with its antipodal points identified), while in space $\R^3$ their relative angles extend to the (half-)sphere. While the circle admits any holonomy or twist $e^{i\theta}$ (thus \emph{any}ons) the additional topological constraints of the sphere reduce these possibilities to either the trivial twist (antipodal symmetry, bosons) or the signed twist (antipodal antisymmetry, fermions).

    \begin{Theorem}[Hardy inequality {\cite{HofLapTid-08,LunSol-13a}}]
        Let
        $$
            \hat{E}_\theta = \hat{\bp}_1^2 + \hat{\bp}_2^2
            = -\nabla_{\bx_1}^2 -\nabla_{\bx_2}^2
        $$
        denote the kinetic energy operator for two particles in $\R^d$,
        $d \in \{2,3\}$, acting on wave functions $\Psi \in L^2(\R^{2d})$ subject to the exchange condition
        $$
            \Psi(\bx_2,\bx_1) = e^{i\theta}\Psi(\bx_1,\bx_2).
        $$
        For $d=2$ we interpret this on the covering space as a continuous simple counterclockwise exchange
        and then allow $\theta \in [0,2\pi)$ (anyons),
        while for $d=3$ we can only have $\theta=0$ (bosons) or $\theta=\pi$ (fermions).
        Then
        $$
            \hat{E}_\theta \ge \left( (d-2)^2/4 + (d-1) \min_{n \in 2\Z} |n+\theta/\pi|^2 \right) \frac{2}{|\bx_1-\bx_2|^2}  
        $$
        in the sense of quadratic forms.
    \end{Theorem}

    It follows that for states $\Psi$ with finite kinetic energy there is if $d=2$ and $\theta \neq 0$ an
    exclusion at the diagonal set $\{\bx_1=\bx_2\}$, an extended probability vortex
    $$|\Psi(\bx_1,\bx_2)|^2 \to 0, \quad \text{as} \ \ |\bx_1-\bx_2| \to 0,$$
    and thus an {\bf exclusion principle} for anyons.
    For particles in space, $d=3$, the repulsive potential on the r.h.s.\ is not strong enough to enforce a probability vortex, and there is even a first term in the constant, which is a global feature. 
    The second term depends on the exchange statistics $\theta$ and persists also in stronger, \emph{local} formulations of the Hardy inequality \cite{LunSol-13a,LarLun-18,LunQva-20} ---
    then in fact strong enough to treat a many-body gas of particles and deduce a {\bf degeneracy pressure} in the gas due to the statistics \cite{LunSol-13b,LunSei-18}.
    It is this pressure that is responsible for the observed {\bf stability} of fermionic matter
    \cite{DysLen-67,LieThi-75}.
    For anyons its precise dependence on $\theta$ is still unknown, and is complicated by the interference of additional particles in the exchange holonomies, due to the uncertainty principle.
    Further, the even more general possibility of {\bf non-abelian anyons} \cite{GolMenSha-85}, for which the fibers are of higher dimension and the twists $T$ are unitary matrices, raise even more questions, e.g.\ concerning statistics transmutability and the existence of global sections \cite{LunQva-20}.
    We refer to \cite{Lundholm-23} for a recent overview of the mathematical physics of the anyon gas, and to \cite{Lundholm-19} for mathematical generalities of uncertainty and exclusion.

	\section{Non flatness, loop quantum gravity inconsistency in multidisciplinary applications}

     In the above example concerning quantum statistics our bundles were locally flat, while twists around certain diagonals of the space gave rise to global curvature.
    One may also consider locally nontrivial holonomies and thus local curvature, which we illustrate in this section concerning quantum gravity and decision theory.

	\subsection{Loop quantum gravity discretization of a connection}
 
	\paragraph*{\bf Presentation of two discretizations of connections and loop quantum gravity.}
	Let $\pi : P \rightarrow M$ be a principal bundle of connected Riemannian base, with structure group $G,$ equipped with a prescribed triangulation or cubification $\tau.$ The canonical maps induced by $\pi$ on relevant objects will be also noted by $\pi$ in the sequel when it carries no ambiguity. We note by $\mathcal{C}$ the space of connections on $P.$ Let $\theta \in \mathcal{C}.$
	
	Let us first assume that $P = M \times G$ is trivial. Then  $\mathcal{C} \sim \Omega^1(M,\mathfrak{g}).$ In this setting, the discretization of $\theta$ is often understood as a generalization of the finite elements method, that performs continuous approximations of an initial smooth function, that are affine at the interior of the subdivision of $M$ defined by the triangulation or the cubification. This approach can be found in Whitney's book \cite{Whit57} and produces $H^1\cap C^0$-convergence at the continuum limit (by refining triangulations); see e.g. \cite{AZ1990}. However, this approach produces gauge-invariance breaking, among other technical difficulties, especially when $G$ is not abelian. Partially for this reason, this discretization does not extend straightway to non-trivial principal bundles. 
	
	Another proposition for a discretization is proposed in the context of loop quantum gravity. The first key heuristic idea relies on the fact that $D_{0} \exp = \mathrm{Id}$ on the Lie group $G,$ which implies that, for ``short'' paths and on a prescribed local trivialization $D \times G$ of $P$ over a star-shaped domain $D,$ $l(\gamma) \theta_{\partial_t \gamma(0)} \sim \exp \theta_{\partial_t\gamma(t)}$ where the exponential is the exponential of smooth paths from $C^\infty(\R,\mathfrak{g})$ to $C^\infty(\R, G)$ and $l(\gamma)$ is the length of $\gamma.$ This approach is highly dependent on the regularity and on the estimates of $\theta,$ and on the local trivialization. However, this is the way how loop quantum gravity (again, heuristically) produces a discretization of a connection, defining ``holonomies along the 1-vertices" of a prescribed triangulation. Without more assumptions on gauges and basepoints,  this approach is ill-defined, and only the holonomy of a loop, which corresponds to the notion of curvature, is (almost up to conjugation) rigorous. 
	\paragraph*{\bf Technical presentation} We follow here \cite{Ma2018-2} for a proposal of rigorous definition of discretized connection and curvature for loop quantum gravity.  The nodes of this triangulation or cubification are assumed indexed by $\mathbb{N},$ noted by $(s_n)_\mathbb{N}$ (the manifold $M$ can be non compact). Recall that, for a fixed index $i_1,\ldots,i_n,$ $\mathrm{St}(s_{i_1},\ldots,s_{i_n})$ is the domain described by the simplices or the cubes with nodes $s_i.$ 
 Fixing $s_0$ as a base point  and $p_0 \in \pi^{-1}(s_0),$ 
\begin{enumerate}
	\item Let $j = \min \left\{ i \in \N^* | s_i \in \mathrm{St}(s_0) \right\}.$ We define $g_{0,j}=1$  and $p_j$ the endpoint of the horizontal path over $[s_0,s_j]$ with starting point $p_0.$ Let $I_2 = \{0;j\}.$
	\item Assume that $I_n$ exists, and that, 
	\begin{itemize}
		\item For each $i \in I_n, $ we have constructed $p_i \in \pi^{-1}(s_i)$ and 
		\item For each $(i,j) \in I_n^2,$ with $i<j$, $g_{i,j}$ is the holonomy of $[s_i,s_j]$, starting at $p_i,$ i.e.  $p_j. g_{i,j}$ is the endpoint of the horizontal path over $[s_i,s_j]$ with starting point $p_i.$ 
	\end{itemize}
	Let $j = \min \left\{ i \in \N \setminus I_n | s_i \in \mathrm{St}(s_k; k \in I_n) \right\}.$ We define 
	\begin{itemize}
		
		\item  $k = \min\{ i \in I_n| s_i \in \mathrm{St}(s_j)\}$ and let $p_j$ the endpoint of the horizontal path over $[s_i,s_j]$ starting at $p_i.$   
		\item for $i \in I_n,$ $g_{i,j}$ is defined such as $p_j.g_{i,j}$ is the endpoint of $[s_i,s_j]$ starting at $p_i.$
		\item $I_{n+1} = I_n \cup \{j\}.$
	\end{itemize}
	
\end{enumerate}

The discretization thus describes the holonomy of the connection along the 1-vertices. We have a first sequence $(p_n)_\N$ which stands as a slice of the pull-back $\left((s_n)_\N\right)^* P$ and if $\mathcal{K}_1$ is the 1-skeleton of $\tau,$ the family $(g_{i,j})_{i<j},$ expresses holonomy elements of the connection $\theta$ on the vertices of $\mathcal{K}_1.$ The holonomy of a smooth path $\gamma,$ for a fixed connection, can be approximated by the discretized holonomies computed along a piecewise smooth path along the vertices of the triangulation, close enough to $\gamma.$ 

	\subsection{Pairwise comparisons, decision theory and all that}
	\vskip 12pt
	\noindent
	\paragraph*{\bf Historical presentation of the Analytic Hierarchy Process and pairwise comparisons.}
	The wide popularity of the pairwise comparison methods in the field of multi-criteria decision analysis is mostly due to their simplicity and their adaptability. Since it is  a natural approach to compare two by two the items under control, it is not surprising that the first systematic use of pairwise comparisons is
	attributed to Ramon Llull \cite{6} during the thirteenth-century.
	
	Very probably, people interested in comparisons passed from qualitative to quantitative evaluations for the obvious gain in simplicity of interpretation. The twen\-tieth-century
	precursor of the quantitative use of pairwise comparison was Thurstone \cite{43} and
	the pairwise comparison method has been progressively improved till the groundbreaking work of Saaty \cite{S1977}. In this article Saaty proposed the Analytic Hierarchy Process
	(AHP), a new multiple-criteria decision-making method. Thanks to the popularity of AHP, the pairwise comparison method has become one of the most
	frequently used decision-making techniques. Its numerous applications include most existing fields such as economy (see e.g. \cite{37}), consumer research \cite{14}, management, military science \cite{CW1985,18}, education and science
	(see e.g. \cite{32}), among others.
    In these settings, comparisons by pairs play a crucial role.
	
	By considering difficult situations, various contexts of generalizations have been proposed, including fuzzy pairwise comparisons (see e.g. \cite{BR2022,48}), linearly ordered groups \cite{KSW2016,W2019} and Lie groups that we develop hereafter from \cite{Ma2018-3} as a main reference.
	\vskip 12pt
	\paragraph*{\bf Gauge-theoretic description}
	Let us start with an example. 
	\begin{example}
	Let us consider three currencies A, B and C and let us explain what happens in the corresponding exchange rates.
			
			\begin{tikzcd}
				A \arrow[rd, "e_{A,C}"' ] \arrow[r, "e_{A,B}"] & B \arrow[d, "e_{B,C}"]\\
				& C
			\end{tikzcd}
			
			For almost any existing currency, one can see that:
			\begin{itemize}
				\item The exchange rates $e_{K,L},$ for $(K,L) \in \{A,B,C\}^2,$ truly gives a reciprocal ratio between the currencies, i.e. $ \forall (K,L) \in \{A,B,C\}^2,\  e_{L,K}=e_{K,L}^{-1}.$
				\item a loop that changes $A$ to $B,$ $B$ to $C$ and $C$ to $A$ may give an inconsistency $$e_{A,B}e_{B,C}e_{C,A} \neq e_{AA} = 1.$$ 
		\end{itemize}
	\end{example}
	 
	In this example, what is called inconsistency is a non-trivial holonomy in the loop $[ABC].$ This situation appears very frequently in various contexts: evaluations are performed by pairwise comparisons which enable one to perform a ranking between items, provided pairwise comparisons are consistent, that is, the holonomy of the loop is trivial. 
	 
	The basic situation is the following. Let us consider $n$ items that have to be compared by pairs. The comparison $a_{i,j}$ between the item $i$ and the item $j,$ is encoded by a number in  $\R_+^*,$ and 
	a pairwise comparisons matrix $(a_{i,j})$ is a $n \times n$ matrix with coefficients in $\R_+^*$ such that $\forall i,j, \quad a_{j,i} = a_{i,j}^{-1}.$ This last property implies $a_{i,i}=1.$
	Consistency in a strict sense on the 2-simplex based on the items $i,j,k$ corresponds then to the relation $$a_{i,j}.a_{j,k} = a_{i,k}$$ and one can measure inconsistency by an inconsistency map $ii$ for which 
	$$A \hbox{ is a consistent PC matrix } \Leftrightarrow ii(A) = 0.$$
	
	One example of particular interest for us is Koczkodaj's inconsistency indicator
	introduced in \cite{K1993} and extended in various references since then. 
	
	\begin{remark}
		Another notion of pairwise comparisons matrices changes the multiplicative group $(\R^*_+,.)$ to the additive group $(\R,+)$ with the same (obvious) adaptations of the formulas. These transformations are natural from a group-theoretical viewpoint and in one-to-one correspondence via logarithmic maps.
	\end{remark}

When the pairwise comparisons matrix is consistent,  one gets \textbf{priority vectors}  $(w_i)_{1 \leq i \leq n}$ for which $\forall i,j,\ a_{i,j} = w_i/w_j.$ This construction is exactly the same as the construction of a connection in loop quantum gravity. Following \cite{Ma2018-1,Ma2018-3}, let $n$ be the dimension of the PC matrices under consideration, and let us consider the $n-$simplex
	$\Delta_n$ as a graph, made of $\frac{n(n-1)}{2}$ edges linking $n$ vertices, equipped with a pre-fixed indexation. Then there is a one-to-one and onto correspondence between edges and the positions of the coefficients in the pairwise comparisons matrix. Hence, each PC matrix $(a_{i,j})$ assigns the coefficient $a_{i,j}$ to the (oriented) edge from the $i-$th vertex to the $j-$th vertex. These are the {holonomies} of a $\R_+^*-$connection. 
    The weights can thereby be understood as a global (discretized) gauge.
	
	Minimizing holonomies is a quantum gravity analog of a classical $\R_+^*-$Yang-Mills theory \cite{RV2014,Seng2011}. Indeed, in the quantum gravity approach \cite{RV2014}, the search for 
 flatness of the curvature relies on the minimization of the distance between the loop holonomies and $1,$ which is exactly the basic requirement of reduction of inconsistencies on triads.
	
	Let us summarize the main correspondences that we have highlighted:
	\vskip 12pt
	\begin{tabular}{|c|c|}
		\hline
		& \\
		Discrete Yang-Mills formalism & Pairwise Comparisons (PC) \\
		& \\
		\hline
		connection & PC matrix \\
		& \\
		flat connection & consistent PC matrix \\
		& \\
		curvature = loop holonomy & inconsistency \\
		& \\
		classical Yang-Mills functional & quadratic average of inconsistency \\
		& on triads \\
		& \\
		``sup'' Yang-Mills functional & Koczkodaj's inconsistency indicator\\
		& \\
		\hline
	\end{tabular}

\section{Coherent and impossible perspectives}

    Illustrative analogies between flatness and logical consistency
    lead us to reflect more generally around various notions of perspective.
 
	\subsection{From pairwise comparisons in a group to (coherent) perspectives}
 
	Classically, the comparisons coefficients $a_{i,j}$ are scaling coefficients, that is, $a_{i,j} \in \R_+^*$. 
 However, our observation of the previous section,
 which identifies pairwise comparisons with discretized connections, gives a natural way to define non-abelian analogs to pairwise comparisons. 
	The states $s_j$ have to belong to a more complex state space $S,$ and in order to have pairwise comparisons, a straightforward study shows that we define matrices with coefficients in a group \cite{Ma2018-3}. In this picture, the pairwise comparisons matrix is the matrix of the holonomies along the edges of the simplex of a connection in the loop quantum gravity picture. 
	
	We note by $PC_I(G)$ the set of pairwise comparisons matrices indexed by $I$ and with coefficients in $G.$ 
		We note by $CPC_I(G)$ the set of consistent PC-matrices.
		
		\begin{Definition}
			A (non normalized, non covariant) inconsistency map is a map 
			$$ii : PC_I(G) \rightarrow V,$$
			where $V$ is a normed vector space, and such that $ii(A)=0$ if $A$ is consistent. Moreover, we say that $ii$
			is faithful if $ii(A) = 0$ implies that $A$ is consistent.  
		\end{Definition}

		If we consider a family of states $(s_i)_I$ such that any $s_i$ cannot be a priori compared directly with any other $s_j.$ We then consider a graph $\Gamma_I$ linking the elements which can be compared. For example, in the previous sections, $\Gamma_I$ was the $1-$skeleton of the simplex. For simplicity, we assume that $\Gamma_I$ is a connected graph, and that at most one vertex connects any two states $s_i$ and $s_j.$ By the way, we get a pairwise comparisons matrix $A$ indexed by $I$ with ``holes'' (with virtual $0-$coefficient) when a $1-$vertex does not exist, and for which $a_{j,i}^{-1} = a_{i,j}.$

		In any framework, the loop quantum gravity formulation of the Yang-Mills theory produces an inconsistency map.

	\subsubsection{Pairwise comparisons in cartography and tunnel building}
	Let us now describe two situations where non-abelian groups rise naturally in a description of errors. For these two examples, the group under consideration is a group of orientation preserving, isometric affine transformations $\mathrm{Is}_n$ of a (finite dimensional) affine space $\R^n.$  
	\vskip 6pt
	\noindent
	\underline{Example: cartography in a forest}
	One of the main features in cartography, or during the recovery of an exit path, in a forest (here in a flat land) is the lack of external point where to get a precise indication on the actual position of the observer. Moves and direction changes can only be measured by self-evaluation. This generates numerous, non-compensating errors in the appreciation of the positions during the path. More precisely, 
	\begin{enumerate}
		\item moves along a straight line can be evaluated by a translation, i.e. a vector in $\R^2$, and
		\item changes of direction can be evaluated as rotations, centered at the position of the observer.
	\end{enumerate}
	Gathering these two aspects, a path in the forest can be assimilated to  succession of moves transcribed by elements $a_{1,2},... a_{n-1,n}$ of the group $\mathrm{Is}_2.$  Thus if one can evaluate the $n-th$ position with respect to the initial one, this gives another element $a_{1,n}.$ Very often $$a_{1,n} \neq a_{1,2}a_{2,3}...a_{n-1,n}.$$
	This is exactly a situation of inconsistency where coefficients are in the non-abelian group $\mathrm{Is}_2.$ 
	\vskip 12pt
	\noindent
	\underline{Example: error in tunnel building}
	The situation is the same in tunnel building, where the surveyors need to indicate, at each step of perforation of a tunnel, in which direction one has to correct the next perforation step underground. Each tunnel starting from each side must meet exactly at the end of the process. For the same reasons as in previous example, the (non-abelian) group under consideration here is $\mathrm{Is}_3.$ Currently, the admissible error is in the range of 1 cm per 100 m of tunnel. This error is admissible inconsistency.   
 
	\subsubsection{Pairwise comparisons in perspective}
	Let $[abcd]$ be a 3 simplex (tetrahedron) in the affine space $\R^3.$
	The 0-vertices $\ba,\bb,\bc$ and $\bd$ define barycentric coordinates in $\R^3,$ that is, any point $\bm \in \R^3$ can be identified by projective coordinates $(w_a,w_b,w_c,w_d) \in \mathbb{P}_4(\R)$ defined by the barycentric equation 
	$w_a (\ba-\bm) + w_b (\bb-\bm) + w_c (\bc-\bm) + w_d (\bc-\bm) = \0.$ 
	Therefore, gluing faces of simplexes plays the same role as changes of coordinates in the atlas of a classical manifold. In order to make an intuitive picture, one can imagine 
    that a complex of glued 3-simplexes (a 3-dimensional CW complex) is ``approximated'' by a smooth 3-manifold. 
    It is a standard result of differential topology \cite{Whit44} that a smooth 3-manifold can be embedded in $\R^6$, 
    but there are examples of 3-complexes that cannot be embedded in $\R^3.$ 
	Moreover, 3-dimensional complexes have to be represented in 2-dimensional pictures by perspectives. Let $[\ba\bb\bc\bd]$ be a tetrahedron.
    \begin{wrapfigure}{rh}{0.5\textwidth}
	\centering
	\includegraphics[width=0.5\textwidth]{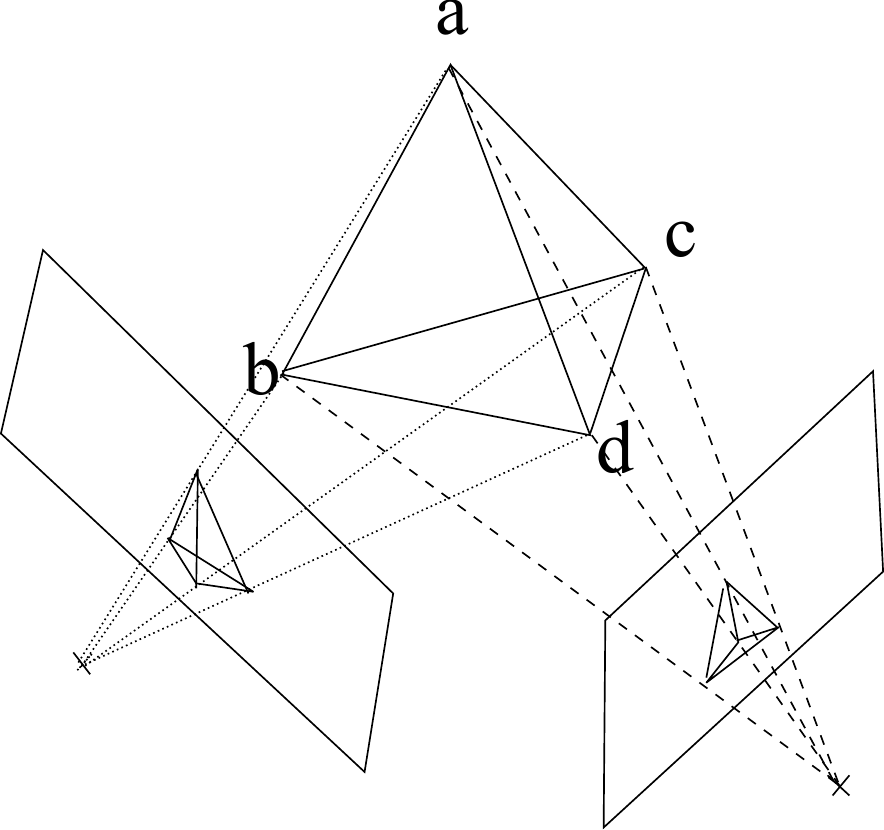}
	\caption[abcd]{Two projections of $[\ba\bb\bc\bd]$}
	\label{fig:proj}
    \end{wrapfigure}
	Let $\omega \in \R^3 \backslash [\ba\bb\bc\bd],$ and let $P_\omega$ be a (projection) plane. In projective perspective, the projection of $\bx \in \R^3 \setminus \{\omega\},$ is $\bx_0 \in P_\omega$ obtained by central projection with respect to $\bomega$.
	Let us consider for simplicity, first, a tetrahedron $[\ba\bb\bc\bd]\subset \R^3,$ and let $\ba_0, \bb_0, \bc_0$ and $\bd_0$ the corresponding projections with respect to $\omega.$
	Any 2-simplex $[\bx\by\bz]$ of $[\ba\bb\bc\bd]$ projects to a 2-simplex $[\bx_0 \by_0 \bz_0] $ in $P_\omega.$ 
	For another choice $(\omega',P_{\omega'}), $ we get other projections $[\bx_0' \by_0' \bz_0'] $ of $[\bx\by\bz].$
	Therefore, if the CW-complex in consideration can be embedded in $\R^3,$ one will be able to get a ``consistently drawn'' picture for almost any chosen point $\omega$ (cf.\ Fig.~\ref{fig:proj}).

	\subsection{Impossible perspectives}
 
\begin{figure}
	\begin{tikzpicture}[scale=0.9]
		\draw [arrows=->,thick] (-2.2,0) -- (2.2,0);
		\draw [arrows=->,thick] (0,-2) -- (0,2);
		\draw [blue,thick] (0,0) ellipse (1.5 and 1.5);
		\node [below right] at (2,0) {$x$};
		\node [below right] at (0,2) {$y$};
	\end{tikzpicture}
	\begin{tikzpicture}[scale=0.8]
		\draw [thick] (0,0) -- (3,0) -- (3,3) -- (0,3) -- (0,0);
		\draw [thick] (0,3) -- (1,4) -- (4,4) -- (4,1) -- (3,0);
		\draw [thick] (3,3) -- (4,4);
		\draw [thick] (0,0) -- (1,1);
		\draw [thick] (1,4) -- (1,1) -- (4,1);
	\end{tikzpicture}
	\scalebox{0.3}{\includegraphics{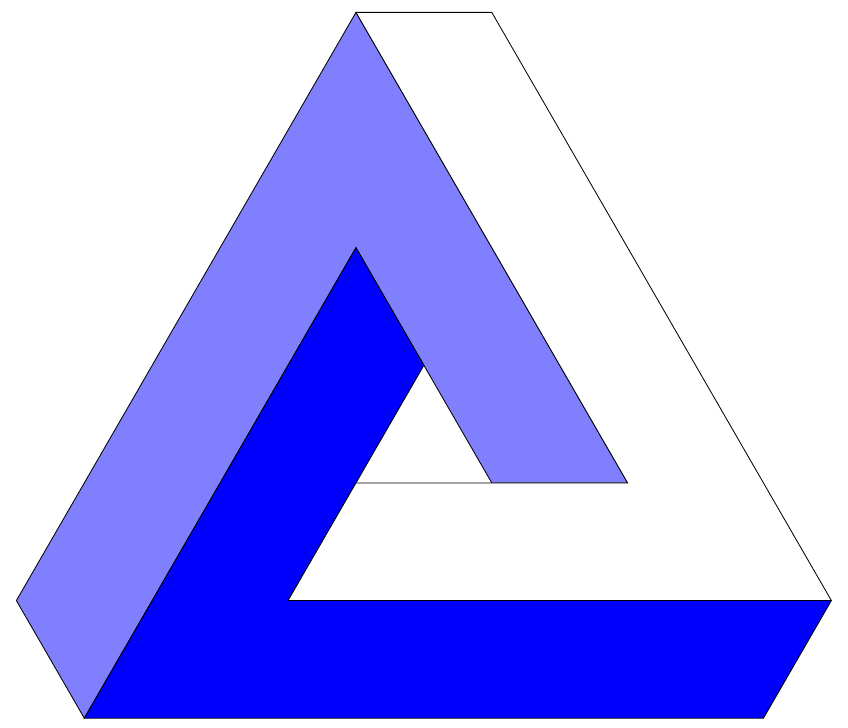}}
    \caption{Choices of perspectives: the circle, the cube and the tribar.}
    \label{fig:perspectives-choice}
\end{figure}
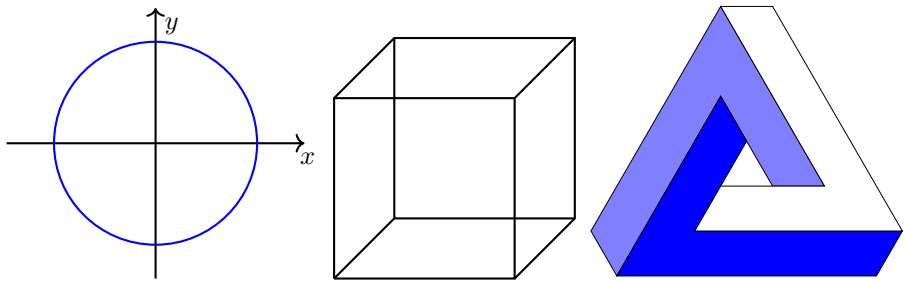

    Upon trying to describe the unit circle defined by the relation $x^2+y^2=1$ using a functional relationship, in which one of the variables is expressed as a function of the other variable, one notes that (cf. Figure~\ref{fig:perspectives-choice}):
    \begin{itemize}
    \item it requires a {\bf choice} of independent and dependent variable: either $x(y)$ or $y(x)$,
    \item given this choice, it also requires a choice of a semicircle (sign of the square root), unless multivalued functions are allowed, and
    \item thus, it is only possible to solve {\bf locally}, and multiple such local descriptions need to be patched together into a global description.\footnote{One may of course also eventually deduce a more clever underlying variable, the angle $\varphi$, from which one obtains a global functional description of the circle.}
    \end{itemize}
    Similarly, the wireframe cube projected to the plane of the paper presents a {\bf choice of perspective}:
    either the cube points out of the plane with its front face towards the lower left, 
    or towards the upper right. These perspectives are not simultaneous, and it is a mental exercise to switch effortlessly between the two.
    Finally, the {\bf Reutersv\"ard--Penrose impossible tribar} presents 
    ``impossibility in its purest form''. Locally, there are three consistent perspectives, involving maximally two corners of the tribar, and presenting us with a choice of which of the corners to exclude from consideration, however, despite their local compatibility, the totality of these local perspectives does not extend to a consistent global perspective. 
    In the words of the Penroses \cite{PP1958}: ``Each individual part is acceptable as a representation of an object normally situated in three-dimensional space; and yet, owing to false connexions of the parts, acceptance of the whole figure on this basis leads to the illusory effect of an impossible structure.''
    Still, the entire figure can easily be drawn and somehow
    makes sense to our minds, 
    and may even stimulate our minds with an aesthetically pleasing concept.

    Although the impossible tribar is often (see, e.g.\ \cite{M2013}) attributed to Lionel and Roger Penrose \cite{PP1958}, who conceived of it independently in 1956 and were inspired by other types of impossible figures drawn by Escher \cite{E1954}, it was first introduced already in 1934 by Oscar Reutersv\"ard, the ``father of the impossible figures'' \cite{O1934}.
    Aiming to set the record straight, Mortensen writes \cite{M2022}:
    ``Their appearance in the art of drawing implies the discovery of an unknown field of some kind of reality. But does that mean an enlarging of `nature’s reality'? Or are the impossible figures simply representations of technical inventions; that is, artefacts? In that case, as Reutersv\"ard himself usually puts it, it is the reality of art that has been enriched by an innovation.''
    Further, the main philosophical message here is
    ``that the human imagination escapes not only the possible, but even the logically consistent; that humans play with contradictory concepts without intellectual collapse.''
    
	\section{Randomness in measurements: decision in non-reciprocal rankings} 

    We now move on to introduce elements of randomness in measurement, 
    and find another instance of twisting of probability in quantum mechanics.
    
	\subsection{Random pairwise comparisons in everyday's life} 
	Let $G$ be a group. We now consider coefficients of matrices in $M_n(G)$ as random variables in a sample probability space $\Omega$ with values in $M_n(G)$ and we call it random pairwise comparisons matrix. The marginal laws on each coefficient $a_{i,j}$ of a random pairwise comparisons matrix $(a_{i,j})_{(i,j)\in \N_n^2}$ determine the probability law of each coefficient. 
	For the sake of motivation and in order to precise accessible situations that are relevant to this picture, using the cases $G= \R$ or $\R_+^*$, we precise that typically the coefficients $a_{i,j}$ follow the following laws:
	\begin{itemize}
		\item a Gaussian law on the \underline{multiplicative} Lie group $\R_+^*$ if the evaluations are honestly made by experts, and
		\item a uniform law with respect to an interval on $\R_+^*$, with respect to the \underline{additive} structure, if the errors are made by human mind corrections or cut-off approximations made with respect to the limits of the measurement procedures and without real link to the problem.
	\end{itemize}
	These two extreme cases are simple consequences of the celebrated Benford's law \cite{B1938} and they are actually used in refined ways for declaration checks, but the law of the random pairwise comparisons matrix $A$ can be anything as well as the laws of the coefficients $a_{i,j}.$
	
	\begin{remark}
		We have here to remark that, by randomness, the use of non-reciprocal pairwise comparisons become natural, even in ``honest'' evaluations due to the natural approximation of the scores as explained before. Manipulated evaluations, due e.g. to a corruption process, are other motivations to consider randomness.
	\end{remark}
 
    \subsection{Contextuality: dealing consistently with twisted perspectives}
	
\mbox{}\\
A clever way to deal logically consistently with twisted perspectives is by means of {\bf ``contextuality''}, namely, we separate different collections of observables that are jointly commensurable into distinct measurement contexts. Each such context thus represents a logically consistent scenario (or `reality') in which all the associated observables can have simultaneous and thus well-defined values. Switching the context corresponds to replacing some of the observables, thus making a new {\bf choice} of the ones that are to be considered, and we are interested in models in which there is a reasonable \emph{local} compatibility between contexts in order to allow for consistent context switching.
We follow Abramsky et al. \cite{ABM2017,ABKLM2015} in this exposition.

Consider a quantum system composed of two subsystems: Alice (A) and Bob (B). Alice has access to two potentially incommensurate observables, $a$ resp. $\alpha$, while Bob has access to the potentially incommensurate $b$ resp. $\beta$.
Each of A and B can choose which one of their respective observables to measure (or know the value of), and their joint choices may thus be organized into four {\bf maximal contexts}:
$$\cM := \bigl\{\{\dred{a},\dblue{b}\},\{\dred{a},\dblue{\beta}\},\{\dred{\alpha},\dblue{b}\},\{\dred{\alpha},\dblue{\beta}\}\bigr\} \subseteq \EuScript{P}(M),$$
where $M := \{\dred{a},\dred{\alpha},\dblue{b},\dblue{\beta}\}$
is the complete set of observables.
The singleton subsets $\{\dred{a}\}$, $\{\dred{\alpha}\}$, $\{\dblue{b}\}$, $\{\dblue{\beta}\}$
and the empty set are (trivial) {\bf subcontexts} of the above, and arise as intersections $C \cap C'$ of $C,C' \in \cM$.
For simplicity we consider only dichotomic observables, i.e.\ the outcomes of any measurements are binary, $O := \{0,1\}$, such as the sign of spin of an electron, or the polarization of a photon, along a given direction in space.

\begin{figure}
	\begin{tikzpicture}[scale=0.7]
		\draw (0,0) -- (2,-0.5) -- (3,0.5) -- (1,1) -- (0,0);
		\draw [dotted] (0,0) -- (0,3);
		\draw [dotted] (2,-0.5) -- (2,2.5);
		\draw [dotted] (3,0.5) -- (3,3.5);
		\draw [dotted] (1,1) -- (1,4);
		\draw [thick] (0,2) -- (2,1.5) -- (3,2.5) -- (1,3) -- (0,2);
		\draw [thick] (0,3) -- (2,2.5) -- (3,3.5) -- (1,4) -- (0,3);
		\draw [fill,red] (0,0) circle(0.05);
		\draw [fill,blue] (2,-0.5) circle(0.05);
		\draw [fill,red] (3,0.5) circle(0.05);
		\draw [fill,blue] (1,1) circle(0.05);
		\draw [fill] (0,2) circle(0.05);
		\draw [fill] (0,3) circle(0.05);
		\draw [fill] (2,1.5) circle(0.05);
		\draw [fill] (2,2.5) circle(0.05);
		\draw [fill] (3,2.5) circle(0.05);
		\draw [fill] (3,3.5) circle(0.05);
		\draw [fill] (1,3) circle(0.05);
		\draw [fill] (1,4) circle(0.05);
		\node [left] at (0,0) {$\dred{a}$};
		\node [right] at (2,-0.5) {$\dblue{b}$};
		\node [right] at (3,0.5) {$\dred{\alpha}$};
		\node [left] at (1,1) {$\dblue{\beta}$};
		\node [left] at (0,2) {$0$};
		\node [left] at (0,3) {$1$};
		\node [above left] at (-2,0) {%
		$\begin{array}{c|cccc}
			\dred{A} \ \dblue{B} & (0,0) & (0,1) & (1,0) & (1,1) \\
			\hline
			(\dred{a},\dblue{b}) & 1/2 & 0 & 0 & 1/2 \\
			(\dred{a},\dblue{\beta}) & 1/2 & 0 & 0 & 1/2 \\
			(\dred{\alpha},\dblue{b}) & 1/2 & 0 & 0 & 1/2 \\
			(\dred{\alpha},\dblue{\beta}) & 1/2 & 0 & 0 & 1/2
            \end{array}$};
	\end{tikzpicture}
	\begin{tikzpicture}[scale=0.7]
		\draw (0,0) -- (2,-0.5) -- (3,0.5) -- (1,1) -- (0,0);
		\draw [dotted] (0,0) -- (0,3);
		\draw [dotted] (2,-0.5) -- (2,2.5);
		\draw [dotted] (3,0.5) -- (3,3.5);
		\draw [dotted] (1,1) -- (1,4);
		\draw [thick] (0,2) -- (2,1.5) -- (3,2.5) -- (1,4) -- (0,3);
		\draw [thick] (0,3) -- (2,2.5) -- (3,3.5) -- (1,3) -- (0,2);
		\draw [fill,red] (0,0) circle(0.05);
		\draw [fill,blue] (2,-0.5) circle(0.05);
		\draw [fill,red] (3,0.5) circle(0.05);
		\draw [fill,blue] (1,1) circle(0.05);
		\draw [fill] (0,2) circle(0.05);
		\draw [fill] (0,3) circle(0.05);
		\draw [fill] (2,1.5) circle(0.05);
		\draw [fill] (2,2.5) circle(0.05);
		\draw [fill] (3,2.5) circle(0.05);
		\draw [fill] (3,3.5) circle(0.05);
		\draw [fill] (1,3) circle(0.05);
		\draw [fill] (1,4) circle(0.05);
		\node [left] at (0,0) {$\dred{a}$};
		\node [right] at (2,-0.5) {$\dblue{b}$};
		\node [right] at (3,0.5) {$\dred{\alpha}$};
		\node [left] at (1,1) {$\dblue{\beta}$};
		\node [left] at (0,2) {$0$};
		\node [left] at (0,3) {$1$};
		\node [above left] at (-2,0) {%
		$\begin{array}{c|cccc}
			\dred{A} \ \dblue{B} & (0,0) & (0,1) & (1,0) & (1,1) \\
			\hline
			(\dred{a},\dblue{b}) & 1/2 & 0 & 0 & 1/2 \\
			(\dred{a},\dblue{\beta}) & 1/2 & 0 & 0 & 1/2 \\
			(\dred{\alpha},\dblue{b}) & 1/2 & 0 & 0 & 1/2 \\
			(\dred{\alpha},\dblue{\beta}) & 0 & 1/2 & 1/2 & 0
            \end{array}$};
	\end{tikzpicture}
 	\begin{tikzpicture}[scale=0.7]
		\draw (0,0) -- (2,-0.5) -- (3,0.5) -- (1,1) -- (0,0);
		\draw [dotted] (0,0) -- (0,3);
		\draw [dotted] (2,-0.5) -- (2,2.5);
		\draw [dotted] (3,0.5) -- (3,3.5);
		\draw [dotted] (1,1) -- (1,4);
		\draw [black!30] (2,1.5) -- (3,3.5) -- (1,4) -- (0,2);
		\draw [black!30] (2,2.5) -- (3,2.5) -- (1,3) -- (0,3);
		\draw (2,1.5) -- (3,2.5) -- (1,4) -- (0,3);
		\draw (2,2.5) -- (3,3.5) -- (1,3) -- (0,2);
		\draw [thick] (0,2) -- (2,1.5);
		\draw [thick] (0,3) -- (2,2.5);
		\draw [fill,red] (0,0) circle(0.05);
		\draw [fill,blue] (2,-0.5) circle(0.05);
		\draw [fill,red] (3,0.5) circle(0.05);
		\draw [fill,blue] (1,1) circle(0.05);
		\draw [fill] (0,2) circle(0.05);
		\draw [fill] (0,3) circle(0.05);
		\draw [fill] (2,1.5) circle(0.05);
		\draw [fill] (2,2.5) circle(0.05);
		\draw [fill] (3,2.5) circle(0.05);
		\draw [fill] (3,3.5) circle(0.05);
		\draw [fill] (1,3) circle(0.05);
		\draw [fill] (1,4) circle(0.05);
		\node [left] at (0,0) {$\dred{a}$};
		\node [right] at (2,-0.5) {$\dblue{b}$};
		\node [right] at (3,0.5) {$\dred{\alpha}$};
		\node [left] at (1,1) {$\dblue{\beta}$};
		\node [left] at (0,2) {$0$};
		\node [left] at (0,3) {$1$};
		\node [above left] at (-2,0) {%
		$\begin{array}{c|cccc}
			\dred{A} \ \dblue{B} & (0,0) & (0,1) & (1,0) & (1,1) \\
			\hline
			(\dred{a},\dblue{b}) & 1/2 & 0 & 0 & 1/2 \\
			(\dred{a},\dblue{\beta}) & 3/8 & 1/8 & 1/8 & 3/8 \\
			(\dred{\alpha},\dblue{b}) & 3/8 & 1/8 & 1/8 & 3/8 \\
			(\dred{\alpha},\dblue{\beta}) & 1/8 & 3/8 & 3/8 & 1/8
            \end{array}$};
	\end{tikzpicture}
    \caption{Empirical models represented by probability tables and corresponding bundle diagrams (cf.\ \cite{ABKLM2015,EM2018}). 
    Top: a non-contextual hidden-variable model. Middle: the strongly contextual PR box. 
    Bottom: the weakly contextual CHSH 
    model.}
    \label{fig:empirical-models}
\end{figure}
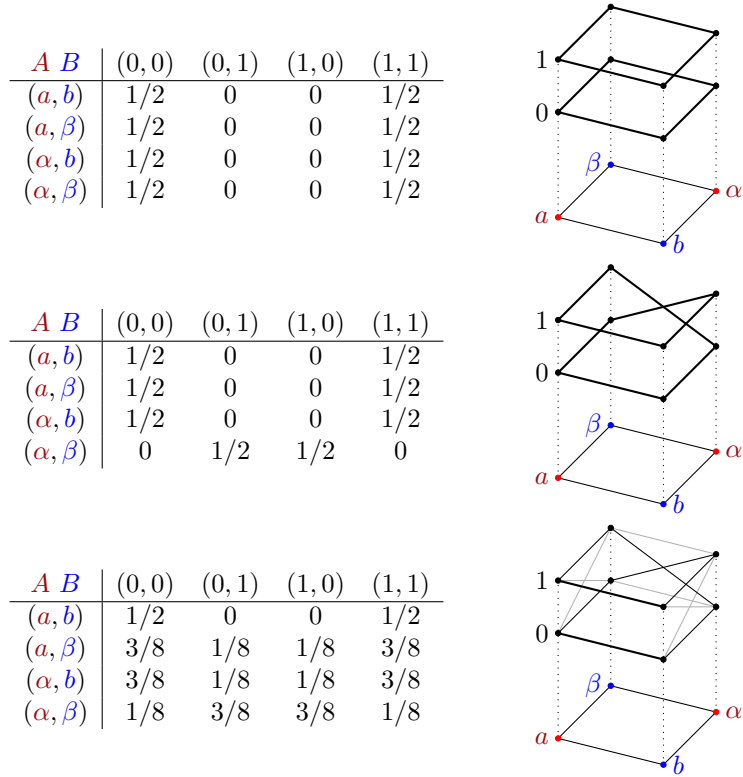

An {\bf empirical model} $(\cM,\bbP)$ consists of an assignment to each context $C \in \cM$ of a probability distribution $\bbP_C$ over the outcomes $C \to O$ of measurements of observables in $C$, i.e.
$$
C \mapsto \left(\bbP_C\colon O^C \to [0,1]\right),
\quad \text{s.t.} \quad
\sum_{s \in O^C} \bbP_C(s) = 1,
$$
and furthermore,
such that there is {\bf local coherence} between intersecting contexts, i.e. compatibility of all the marginals (this implies ``no-signalling''; cf. below)
$$\bbP_C|_{C \cap C'} = \bbP_{C'}|_{C \cap C'}$$
for all contexts $C,C'$. 
Given a subcontext $D \subseteq C$ and a probability distribution $\bbP_C$, we define its {\bf marginalization} onto $D$ by
$$\bbP_C|_D(t) := \sum_{s \in O^C\,:\,s|_D=t} \bbP_C(s)$$
for any measurement outcome $t \in O^D$.
Local coherence thus defines a unique subdistribution $\bbP_D := \bbP_C|_D$ for any subcontext $D \subseteq C \in \cM$.

Consider for example Figure~\ref{fig:empirical-models}, where three different empirical models have been defined using probability tables over the measurement outcomes for each maximal context, and illustrated using bundle diagrams (cf.\ \cite{ABKLM2015,EM2018}), with fiber $O$ over $M$ represented by vertices and the contexts in $\cM$ and probability assignments represented by elementary links.
The first model is a {\bf hidden-variable} model in which the tables may be explained by a single underlying random variable $\lambda \in \{0,1\}$, with $50\%$ probability that any measurements will result in $0$ resp. $1$, given that we know nothing of $\lambda$. Thus, it defines two parallel probable realities, one in which every consecutive measurement yields zero and one in which they yield one, and as soon as we know which of these realities we reside in then we cannot see the other reality by context switching using any single observable.

In the second model, known as the {\bf Popescu-Rohrlich (PR) box} \cite{PR1994}, there is a {\bf twist}, and even though A and B will again measure strictly correlated values for their observables $a$ and $b$ in the context $\{a,b\}$ (both zeros or both ones, with $50\%$ probability), by thereafter choosing to measure $\alpha$ and then $\beta$ (or $\beta$ and then $\alpha$) they will necessarily find an anti-correlation, and eventually, by measuring $a$ and $b$ and thus switching contexts to arrive in $\{a,b\}$ again, they will end up in the parallel reality in which the pair of values have both flipped. Note the inconsistency of a simultaneous assignment of values to the observables: if A knows that $a=0$, then this requires by the first row of the probability table that B knows $b=0$, but then, if A wants to know $\alpha$ as well and thus chooses to measure it, it is required by the third row that she finds $\alpha=0$, whereby by the fourth row B knows that $\beta=1$, and finally by the second row then A knows that $a=1$, a contradiction.

In the third model, known as the {\bf Clauser-Horne-Shimony-Holt (CHSH) model}, there is still a twist but it is a bit weaker.
We may again follow the logic of the model around a loop of measurement contexts, and no longer find an inconsistency in the value assignments, since there is a nonzero probability that we come back to the same value that we started from. However, we do find an inconsistency in the probability assignments, since for example, given $a=0$ there is by the second row $3/4$ probability that $\beta=0$, and then by the fourth row there is $1/4$ probability that $\alpha=0$. On the other hand, there was a remaining $1/4$ probability that $\beta=1$, and then $3/4$ probability that $\alpha=0$.
The total probability that $\alpha=0$ given that $a=0$ is thus $3/16 + 3/16 = 3/8$.
Here we took the path via knowledge of $\beta$.
In the other path, via knowledge of $b$, we conclude by the first and third rows that if $a=0$ then $b=0$, and the total probability that $\alpha=0$ is $3/4$, a contradiction.

We say that an empirical model $\bbP$ is {\bf non-contextual} if there exists
a global assignment of probabilities of outcomes to all observables, i.e.
$$
	\exists f\colon O^M \to [0,1] \ \ \text{s.t} \ \ f|_C = \bbP_C \ \ \forall \ \text{contexts} \ C.
$$
The first model in Figure~\ref{fig:empirical-models} is non-contextual (the global assignment of probabilities is deduced by the hidden variable $\lambda$) while the other two are {\bf contextual} (i.e. non-non-contextual). The second model exemplifies a stronger form of contextuality than the third, since it is impossible to assign any simultaneous values, not just in probability.
However the third model turns out to be realizable in quantum mechanics using a tensor product and a Bell state of entangled photons \cite{CHSH1969}, while the second model turns out not to be.\footnote{The proof of this is based on \textit{Tsirelson's bound}, which gives an upper limit on the allowable violation of Bell inequalities for tensor product quantum systems \cite{Tsi1980}.}

The bundle diagrams in Figure \ref{fig:empirical-models} suggest that the phenomenon of contextuality is somehow `topological' in nature. This connection to topology can be formalized using the language of sheaf theory \cite{AB2011} \cite{BDE2022}. In this language, the probability distributions $\mathbb{P}_C$ over contexts $C$ are local sections of a presheaf of probability distributions. The fact that contextual distributions cannot be realised as the marginals of a global distribution is then precisely the statement that the local sections of this presheaf cannot be glued to a global one. This opens the door to the use of cohomological obstruction theory in the study of contextuality \cite{ABM2011}.

\section{Game theory: using the impossible to solve the impossible}
 
Contextuality and its special case quantum nonlocality can be used as a resource with which to improve coordination between separated parties. The study of \textit{nonlocal games} aims to clarify how two (or more) players who cannot communicate may nevertheless improve their coordination in a cooperative game by leveraging contextuality.

The idea is to phrase contextual measurements in the language of game theory. Suppose Alice and Bob are two players playing a game against some \textit{referee}, who prompts Alice and Bob with one question each and expects an answer in return. Alice and Bob cannot communicate once the questions are posed. If Alice and Bob manage to synchronize their answers, they win together; if they do not, they lose together. The crux is that different question \textit{pairs} may require different answers, even though the individual question that Alice or Bob gets may be the same.

For example, in the \textbf{CHSH game}, Alice and Bob are each prompted with questions labelled 0 or 1, and asked to return an answer from the set $\{0,1\}$. Given a question pair $(x,y)\in\{0,1\}^2$, an answer pair $(a,b)\in\{0,1\}^2$ wins or loses according to Table \ref{tab:win}. Unless Alice and Bob are both prompted with question 1, they should give the same answer. But if they \textit{are} both prompted with 1, they should give different answers. Importantly, Alice and Bob cannot tell from their question alone which question the other was posed with, and hence which answer to give.

\begin{table}[h]
    \centering
    \begin{tabular}{c|cccc}
        \diagbox{$(x,y)$}{$(a,b)$} & (0,0) & (0,1) & (1,0) & (1,1)  \\
        \hline
        (0,0) & W & L & L & W \\
        (0,1) & W & L & L & W \\
        (1,0) & W & L & L & W \\
        (1,1) & L & W & W & L \\
    \end{tabular}
    \caption{Winning and losing answers in the CHSH game. Each row corresponds to a pair of questions $(x,y)$, and each column a pair of answers $(a,b)$. A combination marked with $W$ wins, and a combination marked with $L$ loses.}
    \label{tab:win}
\end{table}

A \textbf{strategy} for playing such a game can be expressed as a probability distribution $\mathbb{P}(a,b|x,y)$, giving the probability that Alice and Bob answer with $a$ and $b$ respectively, when prompted with the question pair $(x,y)$. The assumption that Alice and Bob may not communicate is built in by requiring the {\bf no-signalling} condition that for any questions $x,x',y,y'$, it holds that
\begin{equation*}
    \sum_b \mathbb{P}(a,b|x,y) = \sum_b \mathbb{P}(a,b|x,y'), \quad \text{and} \quad \sum_a \mathbb{P}(a,b|x,y) = \sum_a \mathbb{P}(a,b|x',y).
\end{equation*}
That is, the total probability that Alice answers $a$ cannot depend on Bob's question $y$, and vice versa.

A \textbf{local} strategy is one where Alice and Bob's answers are independent, so that $\mathbb{P}(a,b|x,y) = \mathbb{P}_A(a|x)\mathbb{P}_B(b|y)$ for some distributions $\mathbb{P}_A$ and $\mathbb{P}_B$. This models the classical scenario, where Alice and Bob cannot coordinate at all after the questions are posed. It can be shown that the best local strategy is only as good as the best \textit{deterministic} strategy, where Alice and Bob just answer deterministically according to some functions $f_A : X\to A$ and $f_B : Y\to B$. One can check by hand that the best such deterministic strategy for the CHSH game is that Alice and Bob simply always answer 0, or always answer 1 (some other, more intricate deterministic strategies do just as well, but none do better). This wins every time the question pair is not $(1,1)$. Assuming that all question pairs are equally likely, this means that the optimal strategy wins $3/4$ of the time. However, not every no-signalling strategy is local. There is some wiggle room for Alice and Bob to \textit{coordinate} without \textit{communicating}.

In particular, if Alice and Bob may choose their answers \textit{contextually}, they can win more often. Compare Table \ref{tab:win} with the tables in Figure \ref{fig:empirical-models}. The non-contextual table at the top is an optimal local (but nondeterministic) strategy. However, if Alice and Bob could play according to the strategy given by the strongly contextual PR box in the middle, they would win \textit{every time!} Unfortunately, the PR box is not realisable in quantum mechanics, but the weakly contextual CHSH table is. It does not win every time, but still outperforms the best local strategy with a win rate of $13/16$ (to be compared with $3/4 = 12/16$) under the assumption that all question pairs are equally likely.

By realising the CHSH table with quantum measurements, Alice and Bob are thus able to leverage the contextuality of quantum mechanics in their favor, without ever communicating directly. The phenomenon that contextual strategies are able to outperform local ones is sometimes called \textbf{``quantum pseudo-telepathy''} \cite{BBT2005}, since it may seem that Alice and Bob `communicate without communicating', but really they are just leveraging quantum contextuality to \textit{coordinate} without communicating.

\section{Free will and quantum computing}

    The CHSH game defined in the previous section may be played using a pair of entangled photons and a choice of angles with which to measure their polarizations \cite{CHSH1969}, earning Clauser a share of the 2022 Nobel prize in physics.
    There is another but related contextual setup which is suitable for such games, encoded in the {\bf Bell--Kochen--Specker paradox}:

    \begin{Theorem}[Kochen--Specker {\cite{KS1967}}, Peres {\cite{P1991}}]
	There exists an explicit, finite set of vectors in $\R^3$
	that cannot be $\{0,1\}$-colored in such a way that both of the following
	conditions hold, simultaneously:
	\begin{enumerate}
	   \item For every orthogonal pair of vectors, at most one is colored $0$.
	   \item For every mutually orthogonal triple of vectors, at least one of them
	(and therefore exactly one) is colored $0$.
	\end{enumerate}
    \end{Theorem}

    The set of vectors can be normalized to the sphere $\S^2$ and the theorem thus concerns orthonormal frames in $\R^3$. 
    The proof goes by contradiction, by construction of an explicit finite subset $M \subseteq \S^2$ (remarkably, Peres' 1991 set makes an appearance in Escher's 1961 engraving ``Waterfall'' \cite{M1993})
    and proving the nonexistence of a coloring of $M$ with the required properties.
    This is yet another instance of non-trivial twisting, and applies to the choice
    of a frame in which to measure the polarization of the electromagnetic field,
    i.e.\ the squares of the orthogonal components of spin operators, which commute, take simultaneous values in $\{0,1\}$, and sum to 2.
    The theorem thus says that a polarization state (``a global simultaneous perspective'') that accommodates for every choice of frame (or only those finitely many from $M$) does not exist, and rather the chosen frame constitutes a context with which a polarization (``a local simultaneous perspective'') can be measured.

    Conway and Kochen used this theorem about contextuality to prove what they termed {\bf ``The Free Will Theorem''} \cite{CK2009}, 
    that, ``if an experimenter can freely choose the directions
    in which to orient his apparatus in a certain measurement, then the
    particle's response (to be pedantic---the universe's response near the 
    particle) is not determined by the entire previous 
    history of the universe''.
    In other words, the outcome of measurement is not simply a function of the input arrangements.

    \medskip
    
    In a final example, we turn to {\bf quantum computing}. 
    Namely, it turns out that the `magic' behind powerful quantum computing is again the clever leveraging of contextuality \cite{HWVE2014}.
    A simple version is:

    \begin{Theorem}[Raussendorf {\cite{R2013}}]
    Let $\bbP$ be a measurement-based quantum computer which deterministically evaluates a vector $\mathbf{o}$ of Boolean functions on an input vector $\mathbf{i}$.
    If $\mathbf{o}(\mathbf{i})$ is nonlinear mod 2 in $\mathbf{i}$, then $\bbP$ is strongly contextual.
    \end{Theorem}

    Various versions involving probabilistic evaluation and weaker notions of contextuality, as well as bounding the success probability of computation in terms of the degree of contextuality and degree of nonlinearity, have been studied \cite{ABM2017}.
    Thus, contextuality can be viewed as a resource with which to overcome certain obstacles to computation and solve complexity problems.
    We propose to consider these findings in the light of our discussion,
    and to contrast the {\bf serial/functional} approach taken in classical computation 
    (typically using electrons, i.e.\ fermions, individualized in a Fermi sea)
    to the {\bf parallel/contextual} approach taken in quantum computation
    (possibly using photons, i.e.\ bosons, unified in coherent and entangled states).
    Some types of problems are better treated in the former approach while others are more suited to the latter,
    and the true power seems to lie in the interface between them 
    (statistics transmutation / anyons?).

\begin{figure}
    \begin{tikzpicture}[scale=0.65]
	\draw [dashed,green!50!black] (0.5,-0.5) -- (2,-1.5);
	\draw [dashed,green!50!black] (2.5,-0.5) -- (3,-1.5);
	\draw [dashed,green!50!black] (4.5,-0.5) -- (4,-1.5);
	\draw [dashed,green!50!black] (8.5,-0.5) -- (7,-1.5);
	\draw [fill,green!20] (0,-0.5) -- (1,-0.5) -- (1,0.5) -- (0,0.5) -- (0,-0.5);
	\draw [fill,green!20] (2,-0.5) -- (3,-0.5) -- (3,0.5) -- (2,0.5) -- (2,-0.5);
	\draw [fill,green!20] (4,-0.5) -- (5,-0.5) -- (5,0.5) -- (4,0.5) -- (4,-0.5);
	\draw [fill,green!20] (8,-0.5) -- (9,-0.5) -- (9,0.5) -- (8,0.5) -- (8,-0.5);
	\draw (0,-0.5) -- (1,-0.5) -- (1,0.5) -- (0,0.5) -- (0,-0.5);
	\draw (2,-0.5) -- (3,-0.5) -- (3,0.5) -- (2,0.5) -- (2,-0.5);
	\draw (4,-0.5) -- (5,-0.5) -- (5,0.5) -- (4,0.5) -- (4,-0.5);
	\draw (8,-0.5) -- (9,-0.5) -- (9,0.5) -- (8,0.5) -- (8,-0.5);
	\draw [thick,->] (-1,0) -- (0,0);
	\draw [thick,->] (1,0) -- (2,0);
	\draw [thick,->] (3,0) -- (4,0);
	\draw [thick,->] (5,0) -- (6,0);
	\draw [thick,->] (7,0) -- (8,0);
	\draw [thick,->] (9,0) -- (10,0);
	\draw [thick,->] (1,-1.5) -- (8.5,-1.5);
	\draw [fill] (2,-1.5) circle(0.05);
	\draw [fill] (3,-1.5) circle(0.05);
	\draw [fill] (4,-1.5) circle(0.05);
	\draw [fill] (7,-1.5) circle(0.05);
	\node at (0.5,0) {$f_1$};
	\node at (2.5,0) {$f_2$};
	\node at (4.5,0) {$f_3$};
	\node at (6.5,0) {$\cdots$};
	\node at (8.5,0) {$f_N$};
	\node [above right] at (7.8,-1.5) {$t$};
    \end{tikzpicture}
    \begin{tikzpicture}[scale=0.65]
	\draw [dashed,blue!50!black] (3,0) -- (4.5,-1);
	\draw [dashed,blue!50!black] (6,0) -- (4.5,-1);
	\draw [dashed,blue!50!black] (4.5,1.5) -- (4.5,-1);
	\draw [fill,blue!20] (3,0) circle(0.5);
	\draw [fill,blue!20] (6,0) circle(0.5);
	\draw [fill,blue!20] (4.5,1.5) circle(0.5);
	\draw (3,0) circle(0.5);
	\draw (6,0) circle(0.5);
	\draw (4.5,1.5) circle(0.5);
	\draw [thick,<->] (3.7,0) -- (5.3,0);
	\draw [thick,<->] (3.5,0.5) -- (4,1);
	\draw [thick,<->] (5,1) -- (5.5,0.5);
	\draw [thick,->] (1,-1) -- (8.5,-1);
	\draw [fill] (4.5,-1) circle(0.1);
	\draw (4.5,-1) circle(0.13);
	\node at (3,0) {$U_1$};
	\node at (6,0) {$U_2$};
	\node at (4.5,1.5) {$U_3$};
	\node [above right] at (7.8,-1) {$t$};
    \end{tikzpicture}
    \caption{Two approaches to computation and time: the serial/functional/fermionic approach of a classical computer and the parallel/contextual/bosonic approach of a quantum computer.}
    \label{fig:computation}
\end{figure}
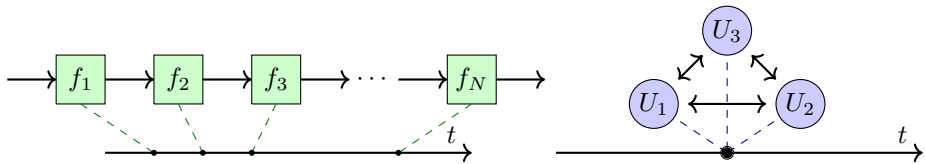

    \medskip

    In conclusion, we may approach to understand quantum mechanics in terms of the impossible tribar in Figure~\ref{fig:perspectives-choice},
    as a succession of concepts:
    \begin{enumerate}
        \item The uncertainty principle, effectively illustrated mainly using 1D concepts. The right angles of the tribar correspond to the projections of knowledge upon context switching using incommensurate observables, and its twisting at each corner yields a $\pi/2$ holonomy on each lap around the loop or M\"obius band of bars.
        \item The exclusion principle, effectively illustrated mainly using 2D concepts, and building upon 1D concepts. The tribar extends conceptually from a 2D projection, and its corners are separated and suspended into their rigid positions due to the twist.
        \item Contextuality, effectively illustrated mainly using 3D concepts, which build upon previous 1D and 2D concepts and their compatibilities. 
    \end{enumerate}
    From this vantage point we could imagine a possible succession to increasingly higher-dimensional concepts, that in order to be discovered and comprehended require a sufficient compactification or condensation of any previous concepts into more manageable components.
In fact, Mortensen writes \cite{M2022}:
``If the flight of the twentieth century from three dimensionality led in fact directly downwards to the plains of two dimensionality, perhaps the art of the coming century will lead art in the opposite direction, up to four-dimensional worlds, where 270$^{\circ}$ triangles and eternal staircases have their place among the normal requisites.''
And Reutersvärd himself envisions:
``The creation of non-Euclidean images is as yet only a novelty in art, that has not come far into use. But undoubtedly the possibility of impossibility will in the future give rise to rich and dizzying visual experiences, when imaginative artists have understood how to further develop the impossible constructions.''

    Further, we may summarize: 
    \begin{itemize}
	\item Uncertainty is about the incommensurability of observables,
	thus expressing the incompatibility or non-simultaneity of perspectives.

	\item Entanglement is about the symmetry or correlation in knowledge,
	thus expressing the compatibility or co-simultaneity of perspectives.
	
	\item Constraints on the observables of a system and uncertainty may together imply certain twisting and vorticity in the correlations of outcomes, and eventually exclusion in probability or other types of tangible correlation or entanglement encoded in local sections.
    An entire arsenal of mathematics, including geometry, topology, analysis, and logic,
    may be required to deduce and analyze such sections and make predictions for physics.
	
	\item Contextuality is about the nonexistence of global sections,
	which then necessitates a choice of a strictly local simultaneous perspective.
        Incompatible perspectives may coexist and even interplay in a cooperative manner to solve seemingly logically impossible problems.
    \end{itemize}
 
	 \section{Outlook}
	 From section 2 to section 7, our six blinds have described their own vision of their restricted problems. In these apparently disconnected features, similar underlying mathematical constructions seem to be present, and to our own perception, these almost constructions are all linked with the problem of measurement, uncertainty, entanglement, and contextuality, 
  which can be 
  summarized as:
  
  \begin{enumerate} 
  \item at an apparent level, an inability of the chosen model, mainly established with our own perception and analysis of the problem under consideration, to fit with the realm. In order to circumvent this problem, many strategies can be developed: enlarge the dimension of the model or invoke hidden variables, or a contrario use symmetries and reduce its complexity, consider an object only as an interacting entity, make the use of probabilities to deal with unattainable measurements, and even consider impossible quantities... all the necessary transformations that make a new model more efficient,
  \item and at a deeper level, the lack of capacity of a human mind to deal with what is hurting its common perception of reality. Indeed, the same way we all know people who have problems to deeply imagine what happens in perspective, the same way a beginner can be puzzled by objects in higher dimensions (the fourth one is perhaps the most popular and the most present in literature), and the same way even an experienced scientist can miss realities hidden to his/her contemplation. 
  Perhaps we must even go so far as to accept that sufficiently complex concepts cannot be rationalized but must be experienced using other modes of perception.
  Our 
  common logical language is itself a human construction built from philosophy and experience. Therefore, all models that can be built by a human mind (and hence, by an artificial intelligence) may in fine express the finitude of its creator. Would the idea of measurement be so artificial?  
  \end{enumerate}

	 	\vskip 12pt
	 	\noindent
	 	\paragraph{Acknowledgements} D.L. and J-P.M. had the chance to be invited, together, to the XL Workshop on Geometric Methods in Physics in Bialowieza 2023 for one plenary lecture each. The deep coincidence of their chosen themes for their lectures led them to gather their efforts for the proceedings, in order to produce a common contribution. This overlap of interest, apparently due to ``a hand of God" who decided their meeting in Bialowieza, has been certainly favorized by the deep insight of the organizers and by their commitment in the constant high quality of the workshop. May they receive our most warm thanks for everything.  
   
   Financial support from the Swedish Research Council 
   (D.L., grant no. 2021-05328, ``Mathematics of anyons and intermediate quantum statistics'')
   and the Center for Interdisciplinary Mathematics at Uppsala University (A.E.)
   is gratefully acknowledged.
   A.E. and D.L. would also like to thank the QMATH group at the University of Copenhagen for an inspiring school on non-local quantum games held in August 2022, leading to new connections and new directions in our research.

\end{document}